\documentclass[journal]{IEEEtran}
\IEEEoverridecommandlockouts
\usepackage{comment}
\usepackage{multicol}
\usepackage{epsfig,graphics,amssymb}
\usepackage{amscd}
\usepackage{amsmath}
\usepackage[normal,bf]{caption2}
\usepackage{enumerate}
\usepackage{theorem}
\usepackage{cite}
\usepackage{epsfig}
\usepackage{verbatim}
\usepackage{color}
\theoremstyle{plain} \theorembodyfont{\itshape}
\theoremheaderfont{\scshape}
\newtheorem{theorem}{Theorem}
\theorembodyfont{\itshape} \theoremheaderfont{\scshape}

\newtheorem{lemma}{Lemma}

\theoremstyle{plain} \theorembodyfont{\itshape}
\theoremheaderfont{\scshape}

\newtheorem{assumption}{Assumption}

\newcommand{\LL}{\rotatebox[origin=c]{180}{$\Lambda$}}
\newcommand{\LLs}{${\small\LL}$}
\begin{document}
\title{Self-Averaging Expectation Propagation}
\author{Burak~\c{C}akmak, Manfred Opper, Bernard~H.~Fleury and Ole Winther
\thanks{Burak~\c{C}akmak and Bernard~H.~Fleury are with the Department of Electronic Systems, Aalborg University, Denmark (e-mail: \{buc, fleury\}@es.aau.dk).}%
\thanks{Manfred Opper is with Department of Artificial Intelligence, Technische Universit\"{a}t Berlin (TUB), Germany  (manfred.opper@tu-berlin.de).} 
\thanks{Ole Winther is with DTU Compute, Danmarks Tekniske Universitet, Denmark  (olwi@dtu.dk).}
}
\maketitle
\begin{abstract}
We investigate the problem of approximate Bayesian inference for a general class of observation models by means of the expectation propagation (EP) framework for large systems under some statistical assumptions. Our approach tries to overcome the numerical bottleneck of EP caused by the inversion of large matrices. Assuming that the measurement matrices are realizations of specific types of  ensembles we use the concept of freeness from random matrix theory to show that the EP cavity variances exhibit an asymptotic self-averaging property. They can be pre-computed using specific generating functions, i.e. the R- and/or S-transforms in free probability, which do not require matrix inversions. Our approach extends the framework of (generalized) approximate message passing -- assumes zero-mean iid entries of the measurement matrix -- to a general class of random matrix ensembles. The generalization is via a simple formulation of the R- and/or S-transforms of the limiting eigenvalue distribution of the Gramian of the measurement matrix.  We demonstrate the performance of our approach on a signal recovery problem of nonlinear compressed sensing and compare it with that of EP.
\end{abstract}

\def\mathlette#1#2{{\mathchoice{\mbox{#1$\displaystyle #2$}}%
                               {\mbox{#1$\textstyle #2$}}%
                               {\mbox{#1$\scriptstyle #2$}}%
                               {\mbox{#1$\scriptscriptstyle #2$}}}}
\newcommand{\matr}[1]{\mathlette{\boldmath}{#1}}
\newcommand{\RR}{\mathbb{R}}
\newcommand{\CC}{\mathbb{C}}
\newcommand{\NN}{\mathbb{N}}
\newcommand{\ZZ}{\mathbb{Z}}
\section{Introduction}
Expectation Propagation \cite{Minka1}, \cite{OW0} (EP) is a typically highly accurate method for approximate probabilistic and Bayesian inference which is applicable to both discrete and continuous random variables as well as hybrid models. Especially Gaussian EP which approximates intractable posterior distributions by multivariate Gaussian densities was found to give excellent approximations not only to posterior marginal moments but also to the free energy i.e. the negative logarithm of the marginal probability density function (pdf) of the observed variables \cite{Opper:2013}. 

Unfortunately, the advantage of Gaussian EP which takes dependencies between variables into account over other methods which are based on simpler approximations with factorizing densities, becomes a problem when the number of random variables is large. This stems from the fact that EP requires frequent matrix inversions related to the update of variance parameters of the Gaussian approximations. This makes a direct application of EP to large systems problematic.

On the other hand there are other approaches to approximate inference which explicitly take advantage of the fact that the number of random variables in the model is large. Central limit theory arguments applied to linear combinations of random variables have been frequently used to facilitate approximate inference \cite{Kabashima}, \cite{Donoha}, \cite{Rangan}. This idea is also very much at the heart of the so-called Thoules-Andersen-Palmer (TAP) approach originally developed in the field of statistical physics and also frequently applied to probabilistic inference \cite{OW0}, \cite{Adatap}, \cite{Kab08}. These approaches lead to Gaussian approximations but with typically simpler parametrization that avoid costly matrix inversions, see also the method of \emph{"diagonal restricted" expectation consistency} \cite{OW5}. This idea has been used e.g.~to develop tractable approximations to (heuristic) loopy belief propagation when the connectivity of the graphical model becomes large \cite{Kabashima}. The so-called approximate message passing (AMP) technique -- originally developed in the context of code division multiple access (CDMA) communication problem\cite{Kabashima} -- has been successfully applied in compressed sensing\cite{Donoha,Rangan,Florent}. This approach relies on statistical assumptions on measurement matrices, assuming that they are random matrices with zero-mean independent and identically distributed (iid) entries. Under this assumption, the variance parameters of the Gaussian random variables become asymptotically non fluctuating and are the same for each variable. This {\em self--averaging} value can be explicitly computed for the simple random matrix ensemble with zero-mean iid entries. 

In this paper we show that under certain statistical assumptions on transformation matrices, the cavity variances computed by EP become self--averaging and can be computed without costly inversions of large matrices. The novel aspect of the our work is to go beyond the simple iid case of AMP algorithm and allow for more general types of dependencies for matrix entries. We develop expressions for the cavity variances of EP in terms of specific generating functions of the matrix statistics via the R- and S--transforms in free probability. Our approach is based on powerful methods of random matrix theory, especially the concept of {\em asymptotic freeness} \cite{Hiai}. The asymptotic limit considered in our paper is that both the number of rows and columns of measurement matrix grow with the fixed aspect ratio of the matrix. This turns out to be different from the standard large data limit frequently considered in statistics, where the posterior distribution becomes highly concentrated around its mode. Moreover, we are also not concerned with convergence issues of specific iterative algorithms for solving EP fixed-point equations in this paper, but concentrate our analysis to the properties of EP fixed points.

Essentially, our technical assumptions require that the system obeys a ``\emph{democratic order}". There is no preferred latent variable in the system, i.e. all latent variables contribute to the data identically in a probabilistic sense. In fact, this assumption is important for the quality of EP per se, because EP approximates the cavity fields by Gaussians, i.e.~it implicitly assumes a Central Limit Theorem to hold. This is again assuming the same kind of ``\emph{everything contributes identically}". 

The paper is organized as follows: Section 2 introduces the system model and discuss how it can obey a `democratic order", i.e. there is no preferred latent variable in the system.  In section~3 we present the EP fixed-point solution of the studied observation model. Section~4 presents a brief summary of the concepts of random matrix theory needed for our approach together with our main mathematical result. Section~5 uses this result to define the self-averaging EP method. In Section~6 we demonstrate the performance of our method on a signal recovery problem of non-linear compressed sensing for two types of random matrices and compare it with EP. Section~7 gives a summary and outlook. Lengthy technical derivations are deferred to the Appendix. 

\subsubsection*{Notations}
The $(n,k)$th entry of an $N\times N$ matrix $\matr X$ is denoted by either $X_{nk}$ or $[\matr X]_{nk}$. We define the normalized trace of $\matr X$ as ${\rm Tr}(\matr X)\triangleq {\rm tr}(\matr X)/N$ and its asymptotic $\phi(\matr X)\triangleq\lim_{N\rightarrow\infty}{\rm Tr}(\matr X)$. $(\cdot)^\dagger$ is the transposition.  The $k$th entry an $K\times 1$ vector $\matr x$ is denoted by either $x_k$ or $[\matr x]_k$. Moreover, $\left<\matr x\right>\triangleq\sum_{k=1}^{K} x_k/K$. For column vectors $\matr x$ and $\matr z$, by abuse of notation we let $(\matr x,\matr z)\equiv(\matr x^\dagger, \matr z^\dagger)^\dagger$. $\mathcal N(\cdot \vert \matr \mu, \matr\Sigma)$ denotes the Gaussian pdf with mean $\matr \mu$ and the covariance matrix $\matr \Sigma$. For random variables $X$ and $Y$, $X\sim Y$ implies that $X$ and $Y$ are identically distributed. For sequences $a_{n}$, $b_{n}$ we imply by $a_n\simeq b_n$ that $a_n-b_n\to 0$ as $n\to \infty$. All large system limits are assumed to hold in the almost sure sense, unless explicitly stated.
\section{The System Model and Its Democratic Order}
We consider a general class of observation models where a $K$-dimensional latent vector $\matr x$ is first linearly transformed with an $N\times K$ dimensional matrix as $\matr z=\matr H\matr x$ and then the vector $\matr z$ is operated according to a function $f(\matr y\vert \matr z)$ where $\matr y$ is the output vector of the system. Here all variables are real-valued. We consider a probabilistic system and refer to $f(\matr y\vert \matr z)$ the likelihood function. Furthermore, we adopt the Bayesian' philosophy and assign a pdf $f(\matr x)$ to the latent variables $\matr x$. Moreover, we consider the typical assumptions that the prior pdf and the likelihood are both separable, i.e. $f(\matr x)=\prod_{i}f_{i}(x_i)$ and $f(\matr y\vert \matr z)= \prod_{j} f_{j}(y_j\vert z_j)$. The probabilistic system can be described by the joint posterior pdf of the latent vector $(\matr x,\matr z)$ as
\begin{equation}
	f(\matr x,\matr z\vert\matr y, \matr H)= \frac{1}{Z}f(\matr x)\delta(\matr z-\matr H\matr x)f(\matr y\vert \matr z) \label{joint}
\end{equation}
where $Z$ denotes a normalization constant. 
\subsection{System obeying a democratic order}
We will restrict the system to obey a ``democratic order". In the system there will be no preferred latent variable in a probabilistic sense. \emph{As a matter of fact this restriction is important for large systems if one needs to accurately understand (and/or control) the system with respect to some macroscopic quantities without making reference to some specific variables. For systems that involve an asymmetry, i.e.~some variables have more or less impact than others, the description should reflect the impact of each variable individually. But, this might be clearly a non-trivial case in the study of large systems.}

Statistical mechanics studies large systems consisting of many elements, e.g. electrons, molecules, etc., which interact with each other \cite{Sethna}. Typically, it uses the Hamiltonian formalism: A so-called Hamiltonian function, i.e. the total energy of the system, which is a sum of the energies of all constituents. Vaguely speaking, it is implicitly assumed that \emph{all (same-type) constituents contribute to the Hamiltonian in a ``democratic fashion"}. There is no preferred constituent.

Our conceptual view is similar to that of statistical mechanics. Specifically, we define a Hamiltonian function for the model \eqref{joint} and restrict the system in way that the latent variables contribute to the Hamiltonian identically in a probabilistic sense. For mathematical convenience we work on a perturbated form of the pdf \eqref{joint} as
\begin{align}
	f_{\tau}(\matr x,\matr z\vert\matr y, \matr H)&=\frac{1}{Z}f(\matr x)\mathcal N(\matr z\vert \matr H\matr x, \tau{\bf I})f(\matr y\vert \matr z)
	\\&= \frac{1}{Z}e^{\mathcal H_{\tau}(\matr x,\matr z\vert \matr y)}\label{joint2}
\end{align}
which yields the pdf \eqref{joint} as $\tau\to 0$. Here $\mathcal H_\tau$ is the Hamiltonian function of the system and it is given by
\begin{align}
	\mathcal H_{\tau}(\matr x,\matr z\vert \matr y)&\triangleq \sum_{i} \ln f_i(x_i)+\sum_j \ln f_{j}(y_j\vert z_j)\nonumber \\&-
\frac{1}{2\tau}(\matr x,\matr z)^\dagger\matr J(\matr x,\matr z)- \frac{N}{2}\ln{2\pi\tau}\label{hamiltonian}
\end{align}
where for convenience we define
\begin{equation}
	\matr J\triangleq\left( \begin{array}{cc}
		\matr H^\dagger \matr H& \matr H^\dagger \\
		\matr H & {\bf I}
	\end{array}\right).
\end{equation}
We now constraint the Hamiltonian function $\mathcal H_{\tau}$ in \eqref{hamiltonian} that have no preferred entry of the latent vectors $\matr x$ and $\matr z$ in a probabilistic sense. Mathematically speaking, let $\matr U$ and $\matr V$ be (uniformly distributed) independent random permutation matrices. Then, we restrict the Hamiltonian to fulfill the symmetry property
\begin{equation}
	\mathcal H_{\tau}(\matr x, \matr z\vert \matr y)\sim \mathcal H_{\tau}(\matr U\matr x, \matr V\matr z\vert \matr V\matr y). \label{democ}
\end{equation}
where we treat $\matr H$ as a random matrix. The democratic order \eqref{democ} can be achieved by fulling the following conditions:
\begin{itemize}
	\item [(i)]$ f_i(x_i)=f_k(x_i)$ for all $i\neq k$;
	\item [(ii)]$ f_j(y_j\vert z_j)=f_l(y_j\vert z_j)$ for all $j\neq l$;
	\item [(iii)] $\matr H\sim\matr U\matr H\matr V$, i.e. the probability distribution of $\matr H$ is invariant under multiplications with independent random permutation matrices from left and right. 
\end{itemize}

The conditions (i) and (ii) are typical assumptions in practice. As regards condition (iii) two points are worth stressing: In case $\matr H$ arises as the result of a description of a physical system, then (iii) is a reasonable assumption if $\matr H$ gives similar weights to the possible interactions between the variables. In case $\matr H$ is specified by design, it is reasonable to restrict the system model such that it fulfills the condition (iii).  Moreover, we here consider a general system model and depending on its specific application some of the restriction above may be useless. For example it might be the case that we solely need 
\begin{equation}
\mathcal H_{\tau}(\matr x, \matr z\vert \matr y)\sim\mathcal H_{\tau}(\matr U\matr x, \matr z\vert\matr y)
\end{equation}
which is fulfilled by the conditions (i) and $\matr H\sim\matr U\matr H$.

\subsection{Haar-Type Eigenvector Matrices}\label{app}
Condition (iii) is still mathematically not convenient to work with in general. We next attempt to explore a convenient random matrix model for $\matr H$ that fulfills (iii). 

We start with the singular value decomposition $\matr H=\matr L \matr S \matr R$ where $\matr L$ is the left eigenvector matrix of $\matr H$, $\matr R$ is its right eigenvector matrix and $\matr S$ is a diagonal matrix whose entries on the diagonal are the singular values of $\matr H$. Condition (iii) holds if (presumedly, if, and only if)  $\matr L$, $\matr S$ and $\matr R$ are independent each other and  $\matr L$ and $\matr R$ are invariant under multiplication with independent random permutation matrices, e.g. $\matr L\sim ~\matr U\matr L$. This implies that \emph{there are no preferred left and right eigenvectors.} We simply restrict this assumption as ``\emph{there is no preferred basis of the left and right eigenvectors}". In other words, $\matr L$ and $\matr R$ are invariant under multiplication with any independent orthogonal matrices -- invariant for short. In summary, we generalize condition (iii) as $\matr H$ is invariant from left and right, i.e. $\matr H\sim\matr U\matr H\matr V$ for orthogonal matrices $\matr U$ and $\matr V$ independent of $\matr H$. Basically, we do not assume a specific distribution for the singular values of $\matr H$ but for the left and/or right eigenvectors. This specific distribution is called ``Haar" \cite{mehta}. This assumption can be also relaxed to some certain Haar-like distribution in the large-system limit \cite{Greg}. \emph{In general, we expect that the analysis of Haar-type random matrices provides an accurate description of systems obeying the democratic order \eqref{democ} i.e. when the contributions of each latent variables to the observation model are statistically identical. }

\section{Expectation Propagation}
This section presents the fixed-points solution of EP approximation for \eqref{joint} and make its connection to the AMP algorithm, see also our previous contribution \cite{SAMPI}. 

For the sake of notational compactness we introduce the compound (column) vector $\matr s\triangleq(\matr x, \matr z)$ and $f(\matr s)\triangleq f(\matr x)f(\matr y\vert \matr z)$. EP approximates the posterior pdf in \eqref{joint} with a Gaussian pdf by substituting the typically non-Gaussian factor $f(\matr s)$ with a separable Gaussian term. Doing so yields the approximation
\begin{equation}
q(\matr s)  \propto  e^{-\frac{1}{2} {\matr s}^\dagger \matr \Lambda \matr s +\matr \gamma^\dagger\matr s} \;
\delta(\matr z-\matr H\matr x)
\label{belief_factor_A}
\end{equation}
where $\matr \Lambda$ is diagonal. The parameters $\matr \gamma$ and $\matr \Lambda$ are computed in an iterative way such that for all $i$ the first and second moments of the marginal $q_i (s_i)$ of the Gaussian pdf  $q(\matr s)$ agree with those of the {\em tilted pdf}, say $\tilde{q}_i (s_i)$, which results from replacing in \eqref{belief_factor_A} the Gaussian factor $e^{-\frac{1}{2}\Lambda_{ii}s_i^2 +\gamma_is_i}$ with the pdf $f_i(s_i)$ and integrating out the remaining variables:
\begin{align}
\tilde{q}_i (s_i) &\propto f_i(s_i) \int
e^{\frac{1}{2}\Lambda_{ii}s_i^2 -\gamma_is_i} q(\matr s)\; {\rm d}\matr s_{\backslash i} \nonumber  \\
&\propto 
f_i(s_i) \exp\left(-\frac{1 }{2} {\small\LL}_{ii}s_i^2 +\rho_is_i\right)   
\label{tilted_dist}.
\end{align}
Here $\{\small \LL_{ii}\}$ are referred to as \emph{cavity} variances. 
For notational convenience we write the diagonal matrices $\matr \Lambda$ and $\matr \LLs$ and the vectors $\matr \gamma$ and $\matr \rho$ in the forms 
\begin{align}
\small\matr \Lambda&=\left(  \begin{array}{cc}
\small\matr \Lambda_{\rm x} & \matr 0 \\
\matr 0 & \small\matr \Lambda_{\rm z}
\end{array}   \right), \quad \matr \gamma=(\matr \gamma_{\rm x},\matr \gamma_{\rm z}) \\
\matr\LLs&=\left(\begin{array}{cc}
\matr \LLs_{\rm x} & \matr 0 \\
\matr 0 & \matr \LLs_{\rm z}
\end{array}   \right),\quad \matr \rho=(\matr \rho_{\rm x},\matr \rho_{\rm z}). \label{diags}
\end{align}
For further convenience we introduce 
\begin{align}
\matr \Sigma_{\rm x}&\triangleq({\small\matr \Lambda_{\rm x}}+\matr H^\dagger {\small\matr \Lambda_{\rm z}}\matr H)^{-1} \label{defa}\\
\matr \mu_{\rm x}&\triangleq\matr \Sigma_{\rm x}(\matr \gamma_{\rm x}+\matr H^\dagger \matr \gamma_{\rm z}).\label{defb} 
\end{align} 
By using standard Gaussian integral identities \cite{MCB} one can show that
\begin{equation}
q(\matr x)=\mathcal N(\matr x\vert \matr \mu_{\rm x}, \matr \Sigma_{\rm x}) 
\end{equation}
and thereby $q(\matr z)=\mathcal N(\matr z\vert \matr H\matr \mu_{\rm x}, \matr H\matr \Sigma_{\rm x}\matr H^\dagger)$. Thus, from the first- and second-order moment consistencies between the pdfs $q_i(s_i)$ and $\tilde q_i(s_i)$, the fixed-point equations of EP for \eqref{joint} are given by the following set of equations
\begin{subequations}
	\label{fix}
	\begin{align}
	\eta_i&=\frac{\gamma_i+\rho_i }{{\small \LL_{ii}+\Lambda_{ii}}}={\begin{cases}
		[\matr \mu_{\rm x}]_{ii} & \quad~~ \:\Lambda_{ii}=[\matr \Lambda_{\rm x}]_{ii} \\
		[\matr H \matr \mu_{\rm x}]_{jj} & \quad~~  \:\Lambda_{ii}=[\matr \Lambda_{\rm z}]_{jj} \end{cases}}
	\label{fix1} \\
	\chi_{i}&=\frac{1}{\small{\Lambda}_{ii}+\small{\LL}_{ii}}={\begin{cases}
		[\matr \Sigma_{\rm x}]_{ii} & \:\Lambda_{ii}=[\matr \Lambda_{\rm x}]_{ii} \\
		[\matr H \matr \Sigma_{\rm x}\matr H^\dagger]_{jj} & \:\Lambda_{ii}=[\matr \Lambda_{\rm z}]_{jj} \end{cases}}\label{fix2}
	\end{align}
\end{subequations}
\emph{where $\eta_i$ and $\chi_i$ are the mean and the variance respectively of the pdf $\tilde q_i(s_i)$ in \eqref{tilted_dist}.} By solving for $\eta_i$ via \eqref{fix} for each $i$ we obtain an approximate of the minimum mean-square error estimator of $s_i$, i.e. $\langle s_i\rangle_{f(\matr s\vert \matr y,\matr H)}\approx \eta_i$.

\subsection{TAP-Like Equations} 
One can introduce numerous fixed-point algorithms that solve \eqref{fix}. In this work we restrict our attention to TAP-like algorithms, e.g. \cite{OW0,Adatap,Kabashima,Donoha,Rangan}. Specifically, we will parameterize \eqref{fix} whose form will be similar to the TAP-like fixed-point equations. This is essentially carried out by bypassing the need for the vector $\matr\gamma$ in the fixed-point equations \eqref{fix}. We keep the forms of \eqref{fix2} and the variables $\matr\eta$ and $\matr\chi$ because these variables are introduced through the variables $\matr \rho, \matr\LLs, \matr \Lambda$ only.  Moreover, from \eqref{fix1} we note that $\matr \mu_{\rm x}=\matr \eta_{\rm x}$ where
$\matr \eta=(\matr \eta_{\rm x},\matr \eta_{\rm z})$. Hence, we solely need to bypass $\matr\gamma$ in representing $\matr \rho$. 
Combining \eqref{defa} and \eqref{defb} we write
\begin{equation}
\matr \gamma_{\rm x}=-\matr H^\dagger \matr \gamma_{\rm z}+(\matr \Lambda_{\rm x}+ \matr H^\dagger \matr \Lambda_{\rm z}\matr H)\matr\mu_{\rm x}.\label{dev1}
\end{equation}
The first equality in \eqref{fix1} for all $\Lambda_{ii}=[\matr \Lambda_{\rm x}]_{ii}$ is equivalent to $\matr \gamma_{\rm x}+\matr \rho_{\rm x}=(\matr \Lambda_{\rm x}+\matr \LLs_{\rm x})\matr\mu_{\rm x}$. Inserting \eqref{dev1} we resolve $\matr \rho_{\rm x}$ as
\begin{align}
\matr \rho_{\rm x}&= \matr H^\dagger \matr \gamma_{\rm z}-(\matr \Lambda_{\rm x}+ \matr H^\dagger \matr \Lambda_{\rm z}\matr H)\matr\mu_{x}+ (\matr \Lambda_{\rm x}+\matr \LLs_{\rm x})\matr\mu_{\rm x} \\
&=\matr H^\dagger (\matr \gamma_{\rm z}- \matr \Lambda_{\rm z}\matr H\matr\mu_{\rm x})+ \matr \LLs_{\rm x}\matr\mu_{\rm x}\\
&=\matr H^\dagger\matr m+\matr \LLs_{\rm x}\matr\mu_{\rm x} \quad \mathrm{with}\quad \matr m\triangleq\matr \gamma_{\rm z}- \matr \Lambda_{\rm z}\matr H\matr\mu_{\rm x}.
\end{align}
The second equality in \eqref{fix1} for all $\Lambda_{ii}=[\matr \Lambda_{\rm z}]_{jj}$ is equivalent to $\matr \gamma_{\rm z}+\matr \rho_{\rm z}=(\matr \Lambda_{\rm z}+\matr \LLs_{\rm z})\matr H\matr\mu_{\rm x}$. From this identity we write 
\begin{align}
\matr m &=(\matr \Lambda_{\rm z}+\matr \LLs_{\rm z})\matr H \matr \mu_{\rm x}-\matr\rho_{\rm z}- \matr \Lambda_{\rm z}\matr H \matr \mu_{\rm x} \\
&=\matr\LLs_{\rm z}\matr H \matr\mu_{\rm x}-\matr \rho_{\rm z}. \label{dev2}
\end{align} 
Combining \eqref{dev1}, \eqref{dev2} with \eqref{fix1} we can recast the latter set of equations as
\begin{subequations}
	\label{tap}
	\begin{align}
	\matr \rho_{\rm z}&= \matr \LLs_{\rm z}\matr H\matr \eta_{\rm x}-\matr m \label{tapl}\\
	\matr m &=\matr \LLs_{\rm z}\matr\eta_{\rm z} -\matr \rho_{\rm z}\\
	\matr \rho_{\rm x}&=\matr \LLs_{\rm x}\matr\eta_{\rm x}+\matr H^\dagger \matr m. \label{tapf}
	\end{align}
\end{subequations}
Here one may argue that the characterization of $\matr \rho_{\rm z}$ provided by \eqref{tap} via the identity $\matr\eta_{\rm z}= \matr H\matr\eta_{\rm x}$ may not be unique. However, for any $i$ $[\matr \eta_{\rm z}]_{i}$ is a strictly increasing function of $[\matr \rho_{\rm z}]_i$ and thereby is bijective. Hence, the characterization is unique. The TAP-like form of the EP fixed-point equation consists of \eqref{tap} and \eqref{fix2}.

\subsection{The AMP Algorithm}\label{AMP} 
The AMP algorithm was originally derived in the context of CDMA \cite{Kabashima}. It re-appeared in the context of compressed sensing \cite{Donoha}. Later on it was generalized for the model \eqref{joint} by \cite{Rangan}. Essentially, it is obtained as a large system limit of heuristic loopy belief propagation where the central limit theorem can be applied when the underlying measurement matrix has independent and zero-mean entries with variance $1/K$. The AMP algorithm proceeds the following iterative equations 
\begin{align}
\matr \rho_{\rm z}(t)&= \matr \LLs_{\rm z}(t)\matr H\matr\eta_{\rm x}(t)-\matr m(t-1)\label{AMP1}\\
\matr m(t) &=\matr \LLs_{\rm z}(t)\matr\eta_{\rm z}(t) -\matr \rho_{\rm z}(t)\\
\matr \rho_{\rm x}(t+1)&=\matr \LLs_{\rm x}(t)\matr\eta_{\rm x}(t)+\matr H^\dagger \matr m(t).\label{AMPS}
\end{align}
Here $\matr\eta(t)\triangleq(\matr\eta_{\rm x}(t),\matr\eta_{\rm z}(t))$ denotes the mean vector of the pdf 
\begin{equation}
\tilde q_t(\matr s)\propto f(\matr s)\left(-\frac{1}{2}\matr s^\dagger{\small\matr \LL}(t) \matr s+ \matr s^\dagger\matr \rho(t)\right)
\label{qtilde}
\end{equation}
where $\matr\rho(t)=(\matr\rho_{\rm x}(t),\matr \rho_{\rm z}(t))$ and the diagonal matrix $\small{\matr \LL}(t)$ is the proper conjugations of the diagonal matrices $\small{\matr \LL}_{\rm x}(t)$ and $\small{\matr \LL}_{\rm z}(t)$, see \eqref{diags}. The cavity variances are updated according to  $\small{\matr \LL}_{\rm x}(t)=v_{\rm x}(t){\bf I}$ and $\small{\matr \LL}_{\rm z}(t)=v_{\rm z}(t){\bf I}$ where 
\begin{align}
v_{\rm x}(t)&=\frac{\alpha(1-v_{\rm z}(t)\langle \matr \chi_{\rm z}(t)\rangle)}{\langle \matr \chi_{\rm x}(t) \rangle}  \\
v_{\rm z}(t)&=\frac{1}{\langle \matr \chi_{\rm x}(t) \rangle}.\label{ampcav}
\end{align}
Here $\alpha\triangleq N/K$ and $\matr\chi(t)\triangleq(\matr\chi_{\rm x}(t),\matr\chi_{\rm z}(t))$ (with $\matr\chi_{\rm x}(t)$ of dimension $K$) denotes the variance of $\tilde q_{t-1}(\matr s)$ .
\subsection{Summary of the Fixed-point Equations for AMP and EP}
AMP and EP share the fixed point equations \eqref{tap}. The fixed-point equations of the cavity variances, differ however. Those of AMP read $\matr \LLs_{\rm x}=v_{\rm x}{\bf I}$ and $\matr\LLs_{\rm z}=v_{\rm z}{\bf I}$ where
\begin{align}
v_{\rm x}&=\frac{\alpha(1-v_{\rm z}\langle \matr \chi_{\rm z}\rangle)}{\langle \matr \chi_{\rm x}\rangle}\\
 v_{\rm z}&=\frac{1}{\langle \matr \chi_{\rm x} \rangle}.
\end{align}
Those of EP are given by \eqref{fix2}. Hence, from an algorithmic point of view, the most expensive operations required in EP are related to the computation of the vector of cavity variances $\matr \LLs$ in terms of $\matr\Lambda$ from \eqref{fix2}. 

We will use the fact that the equations \eqref{fix2} are obtained as the stationary points of the objective function 
\begin{equation}
C_{\matr H}(\matr \Lambda) = \ln\vert\matr \Lambda_{\rm x}+\matr H^\dagger \matr \Lambda_{\rm z}\matr H\vert - \ln \vert \matr \Lambda+\matr \LLs \vert.\label{FE0}
\end{equation}
Under certain asymptotic freeness assumptions on the above-mentioned matrices we will derive an asymptotic limiting expression for the first term in (\ref{FE0}) which depends only on certain random matrix transforms. These involve the limiting distribution of singular values of $\matr H$. The transforms can be pre--computed before iterating the EP algorithm if the random matrix ensemble is explicitly given. In other cases, an approximation based on a finite $\matr H$ can be used. Assuming that the deviation from the asymptotic objective function can also be neglected when we minimize \eqref{FE0} by taking derivatives, we will end up with a type of {\em self-averaging} EP that entirely avoids matrix inversions. 

\section{Random Matrix Theory and Asymptotic Results}
\subsection{Asymptotically Free Random Matrices}
The matrices $\matr X=\matr X^\dagger$ and $\matr Y=\matr Y^\dagger$ are asymptotically free if \cite[Chapter~22]{oxfordSpeicher}
\begin{equation}
\phi \left( \prod_{i=1}^{k}(\matr X^{n_i}-\phi (\matr X^{n_i})\matr {\bf I} ) (\matr Y^{m_i}-\phi (\matr Y^{m_i})\matr {\bf I})\right)=0 \nonumber 
\end{equation}
for all $k\geq 1$ and for all $n_1,m_2,\cdots n_k,m_k\geq 1$.  In a word, the normalized trace of any product of powers of  $\matr X$ and $\matr Y$ centered around their normalized trace vanishes asymptotically. By formally replacing $\phi(\cdot)$ with ${\rm Tr}(\cdot)$ in the definition, we obtain the definition of freeness  in finite dimensions. It is easy to show that any matrix and the identity matrix are free. Yet this is the only known case of free matrices. Asymptotically free random matrices are much more abundant. We now give a well-known example that we will use in the sequel. Let $\matr \Lambda_1$ and $\matr \Lambda_2$ be two real diagonal matrices of dimensions $N\times N$. Furthermore, let the distributions of the entries of $\matr \Lambda_1$ and $\matr \Lambda_2$ be either uniformly bounded or compactly supported in the limit $N\to\infty$ such that in the latter case $\matr \Lambda_1$ and $\matr \Lambda_2$ are assumed to be independent. Moreover, let $\matr U$ be a $N\times N$ Haar orthogonal matrix.  Then $\matr \Lambda_1$ and $\matr U^\dagger \matr \Lambda_1 \matr U$ are asymptotically free. The family of known asymptotically free random matrices is growing every day. For example, in the context of complex matrix ensembles, it has been recently shown that the above asymptotic freeness property still holds when one replaces the transposition with conjugate transposition and the Haar matrix with a randomly permuted Fourier matrix (also called ``fake'' Haar unitary matrix) \cite{Greg}. Hence, it is reasonable to expect that a randomly permuted discrete cosine transform (DCT) matrix behaves asymptotically as a Haar orthogonal matrix, see \cite{Enzo}, so that the asymptotic freeness conditions hold.

\subsection{Additive and Multiplicative Free Convolutions}\label{adfmfc}
Asymptotic freeness of matrices allows us to simplify traces of products of non--commuting matrices. This can be compared to the role of independence for simplifying expectations of products of ordinary random variables
in probability theory.  
Conceptually the R-transform is the counterpart in random matrix theory to the cumulant generating function in classical probability theory. It allows for dealing with the sum of free random matrices. Suppose that $\matr X=\matr X^\dagger$ and $\matr Y=\matr Y^\dagger$ are asymptotically free. Then, 
\begin{equation}
{\rm R}_{\matr X +\matr Y}(\omega)={\rm R}_{\matr X}(\omega)+{\rm R}_{\matr Y}(\omega)
\end{equation}
where ${\rm R}_{(\cdot)}$ is called the R-transform of the LED of the matrix given in the subscript.
For our purpose we define the R-transform for a probability distribution $\rm F$ with support in $[0,\infty)$: We introduce the Stieltjes transform of $\rm F$ as
\begin{equation}
{\rm G}(s)\triangleq\int \frac{{\rm dF}(x)}{s-x}, \quad-\infty<s<0.
\end{equation}
Moreover, let $\chi\triangleq\int x^{-1}{\rm dF}(x)$ with the convention $\chi=\infty$ unless ${\rm F}(0)=0$.
Then, the R-transform of ${\rm F}$ is defined as
\begin{equation}
{\rm R}(\omega)\triangleq {\rm G}^{-1}(\omega)-\omega^{-1}, \quad -\chi<\omega<0 \label{r-transform}
\end{equation}
where ${\rm G}^{-1}$ denotes the composition inverse of $\rm G$. In particular, for a distribution $\rm F_{a}$ with support in $[-a,\infty)$, $a>0$, we can first derive the R-transform of the ``shifted"  distribution ${\rm F}(x)={\rm F}_{a}(x-a)$, then obtain the R-transform of $\rm F_{a}$ via 
\begin{equation}
{\rm R}_{a}(\omega)={\rm R}(\omega)-a
\end{equation}
where ${\rm R}_a$ is the R-transform of ${\rm F}_a$.

Similarly, the S-transform is the counterpart in random matrix theory to the Mellin transform in classical probability theory. It allows for dealing with the product of free random matrices.  Specifically, if $\matr X=\matr X^\dagger$ and $\matr Y=\matr Y^\dagger$ are asymptotically free,
\begin{equation}
{\rm S}_{\matr X \matr Y}(z)={\rm S}_{\matr X}(z){\rm S}_{\matr Y}(z)
\end{equation}
where ${\rm S}_{(\cdot)}$ denotes the S-transform of the LED of the matrix given in the subscript. We define the S-transform of for a probability distribution ${\rm F}$ with support in $[0,\infty)$ such that $\alpha\triangleq 1-{\rm F}(0)$ is non-zero: Let $\rm G$ be the Stieltjes transform of $\rm F$ and define 
\begin{equation}
\Psi(s)\triangleq s^{-1}{\rm G}(s^{-1})-1,\quad -\infty<s<0.
\end{equation}
The {S-transform} of ${\rm F}$ is defined as \cite{Voi92}
\begin{equation}
{\rm S}(z)\triangleq\frac{z+1}{z}\Psi^{-1}(z),\quad -\alpha<z<0.
\label{s-transform}
\end{equation}

Finally, we point out that for a given probability distribution $\rm F$ on the real line, its R-transform and S-transform can be conveniently related to each other via the identity \cite{hager}
\begin{equation}
{\rm R}(z{\rm S}(z)){\rm S}(z)=1={\rm S}(\omega {\rm R}(\omega)){\rm R}(\omega).\label{trans}
\end{equation}
This trivial observation has a key role in our derivation as it allows us to formulate the results that are expressed in terms of R-transform via the S-transform and vice versa. 

\subsection{The Asymptotic Result}
To approximate the term  $\ln\vert\matr \Lambda_{\rm x}+\matr H^\dagger \matr \Lambda_{\rm z}\matr H \vert$ in the objective function $C_{\matr H}(\matr \Lambda)$ in \eqref{FE0}, we employ additive and multiplicative free convolutions. We will assume that $\matr \Lambda_{\rm x}$ and $\matr H^\dagger \matr \Lambda_z \matr H$ have LEDs, are asymptotically free and use the result
\begin{equation}
{\rm R}_{\matr \Lambda_{\rm x}+\matr H^\dagger \matr \Lambda_z \matr H}(\omega)={\rm R}_{\matr \Lambda_{\rm x}}(\omega)+{\rm R}_{\matr H^\dagger \matr \Lambda_z \matr H}(\omega) \label{adf}
\end{equation}
The R-transform of the LED $\matr H^\dagger \matr \Lambda_z \matr H$ is still analytically intractable. If $\matr \Lambda_{\rm z}$ and $\matr H \matr H^\dagger$ are asymptotically free, we can resolve these difficultly by making use of
\begin{equation}
{\rm S}_{\matr \Lambda_z\matr H\matr H^\dagger}(z)={\rm S}_{\matr \Lambda_z}(z){\rm S}_{\matr H\matr H^\dagger}(z)\label{mdf}.
\end{equation}
\begin{assumption}\label{as1} 
	In the large system limit, let the matrices $\matr \Lambda_{\rm x}$, $\matr \Lambda_{\rm z}$ and $\matr H^\dagger \matr H$  have compactly supported LEDs and the LED of $\matr \Lambda_{\rm z}$ has its support in $[0,\infty)$ and the maximum eigenvalue of $\matr H^\dagger \matr H$ has a finite limit. Furthermore, ${\phi}(\matr \Lambda_{\rm x}+\matr H^\dagger \matr \Lambda_z \matr H)^{-1}<{ \phi}(\matr H^\dagger \matr \Lambda_z \matr H)^{-1}$ where by convention $\phi(\matr X^{-1})=\infty$ if $\matr X$ is singular. Moreover, \eqref{mdf} and \eqref{adf} hold.
\end{assumption}
The restriction of non-negativeness of $\matr \Lambda_z$ can be relaxed in the analysis.  However, based on numerical evidence and the so-called Almedia-Thouless (AT) line of stability analysis in Appendix~B we conjecture that in practice any effective solution fulfills this condition when the dimensions are sufficiently large. Typically, $\alpha =N/K<1$ and thereby ${\phi}(\matr H^\dagger \matr \Lambda_z \matr H)^{-1}=\infty$, so that the last but one condition is fulfilled. As regards the final condition two points are worth noting. Firstly, if $\matr H$ is invariant from right and left (see Section~\ref{app}), then \eqref{adf} and \eqref{mdf} always hold provided the matrices in Assumption~\ref{as1} have compactly supported LEDs\cite{Collins-a}. Secondly, depending on the application some of the matrices $\matr \Lambda_{\rm x}$, $\matr \Lambda_{\rm z}$, $\matr H\matr H^\dagger$ might be proportional to the identity matrix. For instance, in the context of (linear or nonlinear) compressed sensing  $\matr H$ might be row-orthogonal, i.e. $\matr H \matr H^\dagger={\bf I}$ and thereby \eqref{mdf} always holds.

\begin{theorem}\label{main}
	Let the matrices $\matr \Lambda_{\rm x}$ , $\matr \Lambda_{\rm z}$ and $\matr H$ fulfill the conditions stated in Assumption~\ref{as1} and $(\matr \Lambda_{\rm x}+\matr H^\dagger \matr \Lambda_z \matr H)$ be positive definite. Then, for sufficiently large $N,K$ there exist positive quantities $\chi_{\rm a}$, $v_{\rm a}$ and $\lambda_{\rm a}$ for ${\rm a}\in \{{\rm x}, {\rm z}\}$ such that 
	\begin{align}
	\ln \vert\matr \Lambda_{\rm x}+\matr H^\dagger \matr \Lambda_z \matr H\vert=\ln\vert\matr \Lambda_{\rm x}+ {v}_{\rm x}\matr{\bf I}\vert+\ln\vert\matr \Lambda_{\rm z}+ {v}_{\rm z}\matr{\bf I}\vert\nonumber \\
	+\ln \vert\lambda_{\rm x}\matr {\bf I}+ \lambda_{\rm z} \matr H^\dagger\matr H \vert+K\ln\chi_{\rm x}+N \ln \chi_{\rm z}+\epsilon
	\label{nice}
	\end{align}
	where $\epsilon=O(1)$ is a bounded function of $N$. The quantities in \eqref{nice} are uniquely characterized by the implicit equations
	\begin{align}
	{v}_{\rm x} &=\lambda_{\rm z}{\rm R}^K_{\matr H^\dagger\matr H}(-\lambda_{\rm z}\chi_{\rm x}) \\
	{v}_{\rm z} &=\lambda_{\rm x}{\rm S}^N_{\matr H\matr H^\dagger}(-\lambda_{\rm z}\chi_{\rm z}) \label{cav1} 
	\end{align}
	where $\chi_{\rm a}={\rm Tr}(\matr \Lambda_{\rm a}+ {v}_{\rm a}\matr{\bf I})^{-1}$ and $\lambda_{\rm a}=\chi_{\rm a}^{-1}-{v}_{\rm a}$ for ${\rm a}\in \{{\rm x}, {\rm z}\}$. Here, ${\rm R}^K_{\matr H^\dagger \matr H}$ and ${\rm S}^N_{\matr H \matr H^\dagger}$ denote the R-transform and S-transform of the \emph{empirical} eigenvalue distribution of the matrices $\matr H^\dagger \matr H$ and $\matr H \matr H^\dagger$, respectively. Moreover, we have $\chi_{\rm x}\simeq {\rm Tr}(\matr \Lambda_{\rm x}+\matr H^\dagger \matr \Lambda_z \matr H)$ and
	 \begin{align}
	 {v}_{\rm x}&\simeq\lambda_{\rm z}{\rm R}_{\matr H^\dagger\matr H}(-\lambda_{\rm z}\chi_{\rm x}) \\
	 {v}_{\rm z}&\simeq \lambda_{\rm x}{\rm S}_{\matr H\matr H^\dagger}(-\lambda_{\rm z}\chi_{\rm z}).\label{cav2} 
	 \end{align}
	 {Proof: See Appendix~A.}
\end{theorem}
\section{The Self-Averaging EP Framework}\label{SEP}
To characterize the cost function \eqref{FE0} with respect to $\{\Lambda_{ii}\}$ we use Theorem~\ref{main} as follows:
\begin{align}
\frac{\partial \ln\vert\matr \Lambda_{\rm x}+\matr H^\dagger \matr \Lambda_{\rm z}\matr H \vert}{\partial \Lambda_{ii}}&=\frac{1}{\Lambda_{ii}+{v}}+\frac{\partial \epsilon}{\partial \Lambda_{ii}}\nonumber \\
&+K\chi_{\rm x} \frac{\partial v_{\rm x}}{\partial \Lambda_{ii}}+
N\chi_{\rm z} \frac{\partial v_{\rm z}}{\partial \Lambda_{ii}}+ K\chi_{\rm x}\frac{\partial \lambda_{\rm x}}{\partial \Lambda_{ii}} \nonumber \\ 
& \underbrace{+N\chi_{\rm z}\frac{\partial \lambda_{\rm z}}{\partial \Lambda_{ii}}+K \frac{1}{\chi_{\rm x}}\frac{\partial \chi_{\rm x}}{\partial \Lambda_{ii}}+N \frac{1}{\chi_{\rm z}}\frac{\partial \chi_{\rm z}}{\partial \Lambda_{ii}}}_{=0}\label{good}    
\end{align}
where by abuse of notation we write ${v}={v}_{\rm x}$ for $\Lambda_{ii}=[\matr \Lambda_{\rm x}]_{ii}$ and ${v}={v}_{\rm z}$ for $\Lambda_{ii}=[\matr \Lambda_{\rm z}]_{jj}$. The fact that the sum of the last six terms vanishes is an immediate consequence of the definition of the variables $\chi_{\rm a}$, $\lambda_{\rm a}$ for ${\rm a}\in \{{\rm x},{\rm z}\}$ in Theorem~\ref{main}. Taking the partial derivative of the left-hand term in \eqref{good} and making use of \eqref{fix2} we obtain
\begin{equation}
\frac{1}{\Lambda_{ii}+ {\small\LL}_{ii}}=\frac{1}{\Lambda_{ii}+{v}}+\frac{\partial \epsilon}{\partial \Lambda_{ii}}. \label{key3}
\end{equation}
An explicit analysis of the derivative of the asymptotic correction term, i.e. $\frac{\partial \epsilon}{\partial \Lambda_{ii}}$ requires an extensive random matrix study. Instead, we consider the following heuristic argument: Firstly we recall result $\chi_{\rm x}\simeq{\rm Tr}(\matr \Lambda_{\rm x}+\matr H^\dagger \matr \Lambda_z \matr H)^{-1}$ of Theorem~\ref{main}. Thereby, we have
\begin{equation}
\sum_{i}\frac{\partial \epsilon}{\partial [\matr \Lambda_{\rm x}]_{ii}}=O(1). \label{B1}
\end{equation}
Then, we consider the implicit assumption that ``\emph{everything contributes in a democratic fashion}", specifically there is no preferred individual term in the sum \eqref{B1}. In doing so we have
\begin{equation}
\frac{\partial \epsilon}{\partial [\matr \Lambda_{\rm x}]_{ii}}= O\left( \frac{1}{N}\right). \label{sym}
\end{equation}
Secondly, from \eqref{sym} and using Theorem~\ref{main} one can easily show that
\begin{equation}
\chi_{\rm z}\simeq{\rm Tr}(\matr H(\matr \Lambda_{\rm x}+\matr H^\dagger \matr \Lambda_z \matr H)^{-1}\matr H^\dagger).
\end{equation}
Thereby, we have
\begin{equation}
\sum_{j}\frac{\partial \epsilon}{\partial [\matr \Lambda_{\rm z}]_{jj}}=O(1).\label{B2}
\end{equation}
Similarly, we assume that  ``\emph{everything contributes in a democratic fashion}", specifically there is no preferred individual term in the sum \eqref{B2}. Doing so we have
\begin{equation}
\frac{\partial \epsilon}{\partial [\matr \Lambda_{\rm z}]_{jj}}=O\left( \frac{1}{N}\right). 
\end{equation}
In summary we conclude that we have ${\small \matr \LL_{\rm x}}\simeq { v}_{\rm x}{\bf I}$ and ${\small\matr \LL_{\rm z}}\simeq{v}_{\rm z}{\bf I}$. This means that 
$\small \matr \LL_{\rm x}$ and $\small\matr \LL_{\rm z}$ are asymptotically self averaging.
\subsection{Summary of Self-averaging EP}
For convenience, we first factorize the \emph{tilted} pdf $\tilde q(\matr s)=\prod_{i} \tilde q_i(s_i)$ in \eqref{tilted_dist} as $\tilde q=\tilde q_{\rm x}\cdot \tilde q_{\rm z}$  where
\begin{align}
\tilde q_{\rm x}(\matr x)&\propto f(\matr x)\exp\left(-\frac{v_{\rm x}}{2}\matr x^\dagger \matr x+\matr x^\dagger\matr \rho_{\rm x}\right)\label{qx} \\
\tilde q_{\rm z}(\matr z)&\propto f(\matr y\vert \matr z)\exp\left(-\frac{v_{\rm z}}{2}\matr z^\dagger \matr z+\matr z^\dagger\matr \rho_{\rm z}\right)\label{qz}.
\end{align}
Moreover, for ${\rm a}\in \{ {\rm x}, {\rm z} \}$ let $\matr \eta_{\rm a}$ and $\matr \chi_{\rm a}$ be the mean and variance vectors of the pdf $q_{\rm a}$, respectively. Thus, we have  $\chi_{\rm a}\simeq \langle \matr\chi_{\rm a}\rangle$ where $\chi_{\rm a}$ is given in Theorem~\ref{main}. Thereby, from \eqref{tap}
the self-averaging EP fixed-point equations can be summarized as
\begin{subequations}
	\label{taps}
	\begin{align}
	\matr \rho_{\rm z}&= v_{\rm z}\matr H\matr \eta_{\rm x}-\matr m \\
	\matr m &=v_{\rm z}\matr\eta_{\rm z} -\matr \rho_{\rm z}\\
	\matr \rho_{\rm x}&=v_{\rm x}\matr\eta_{\rm x}+\matr H^\dagger \matr m\\
	{v}_{\rm x}&=\lambda_{\rm z} {\rm R}_{\matr H^\dagger \matr H}(-\lambda_{\rm z}\langle \matr\chi_{\rm x}\rangle)\label{cavn1}\\
	{v}_{\rm z}&=\lambda_{\rm x} {\rm S}_{\matr H\matr H^\dagger}(-\lambda_{\rm z}\langle \matr\chi_{\rm z}\rangle)\label{cavn2}\\
	\lambda_{\rm x}&=1/\langle \matr\chi_{\rm x}\rangle-{v}_{\rm x}\\
	\lambda_{\rm z}&=1/\langle\matr\chi_{\rm z}\rangle-{v}_{\rm z}.
	\end{align}
\end{subequations}
In addition, one can alternatively consider the characterization of $v_{\rm x}$ in \eqref{cavn1} as (see Appendix~A)
\begin{equation}
v_{\rm x}=\frac{\alpha (1-{v}_{\rm z}\langle \matr\chi_{\rm z}\rangle) }{\langle \matr\chi_{\rm x}\rangle} \label{compact}
\end{equation}  
where $\alpha=N/K$. 

When the analytical expression of either the R-transform or the S-transform in the above expressions is known, while the other is unknown, we can use \eqref{trans} to express the cavity variances as a function of the known transform. For example, by using \eqref{trans} we can write \eqref{cavn2} in the form
\begin{align}
{v}_{\rm z}= \frac{\lambda_{\rm x}}{{\rm R}_{\matr H\matr H^\dagger}(-{v}_{\rm z}(1-{v}_{\rm z}\langle\matr\chi_{\rm z}\rangle)/\lambda_{\rm x})}.\label{sR}
\end{align}

It might be the case that the analytical expressions of both the R-transform and the S-transform are unknown. In fact, the LEDs themselves might be even unknown. In such cases the simplest approach would be to use the R-transform ${\rm R}^K_{\matr H^\dagger \matr H}$ and S-transform  
${\rm S}^N_{\matr H \matr H^\dagger}$ of the  \emph{empirical} eigenvalue distribution of the matrices $\matr H^\dagger \matr H$ and $\matr H \matr H^\dagger$, respectively. Using the definitions of the transforms this would lead to the fixed--point equations 
\begin{equation}
\langle \matr\chi_{\rm a}\rangle=\frac{1}{\lambda_{\rm a}+v_{\rm a}}={\begin{cases}
	{\rm Tr}(\lambda_{\rm x}{\bf I}+\lambda_{\rm z}\matr H^\dagger\matr H)^{-1} & \:{\rm a}={\rm x} \\
	{\rm Tr}(\matr H(\lambda_{\rm x}{\bf I}+\lambda_{\rm z}\matr H^\dagger\matr H)^{-1}\matr H^\dagger)& \:{\rm a}={\rm z} \end{cases}}. \label{finite1}
\end{equation}
We can iteratively solve these fixed-point equations without the need for a matrix inversion. The singular values of $\matr H$, which are required in the iterations, are pre--computed. Finally, it is also important to note that the resulting solutions for $\lambda_{\rm x}$ and $\lambda_{\rm z}$ in self-averaging EP should be positive. Otherwise they may lead to an instability of the algorithm and yield incorrect solutions for the cavity variances. 

\section{Numerical Results}
To illustrate our analysis we consider a signal recovery problem from one-bit compressed sensing, see \cite{onebit} and the references therein. Specifically, the signal model reads 
\begin{equation}
\matr y={\rm sign}(\matr H\matr x)
\end{equation}
where $\matr x$ has entries drawn independently from a standard Bernoulli-Gaussian, specifically, with a prior pdf of the {\em spike and slab} form 
\begin{equation}
f(\matr x)=(1-\rho)\delta(\matr x)+\rho \mathcal N(\matr x\vert \matr 0, \tau{\bf I}).
\end{equation}
We restrict ourselves to simulated data. We consider two classical random matrix models in compressed sensing:
\begin{itemize}
	\item [(i)] the entries of $\matr H\in \RR^{N\times K}$ are iid Gaussian with zero mean and variance $1/K$;
	\item [(ii)] a random row-orthogonal model \cite{Candes, Mikko, Tulino13}, namely where the rows of $\matr H$  are drawn from a randomly permuted DCT matrix. Specifically, $\matr H= \matr P (\matr P_{\pi} \matr \Psi \matr P_{\pi}^\dagger)$ where $\matr P \in \{0,1\}^{N\times K}$ with ones on the diagonal and zeros elsewhere, $\matr P_{\pi}$ is a $K\times K$ permutation matrix associated with the permutation $\pi$ which is drawn uniformly from the set of permutations $(1,\cdots, K)\to (1,\cdots, K)$ and $\matr \Psi$ is the $K\times K$ DCT matrix. 
\end{itemize}

\begin{figure*}
	\centering
	\epsfig{file=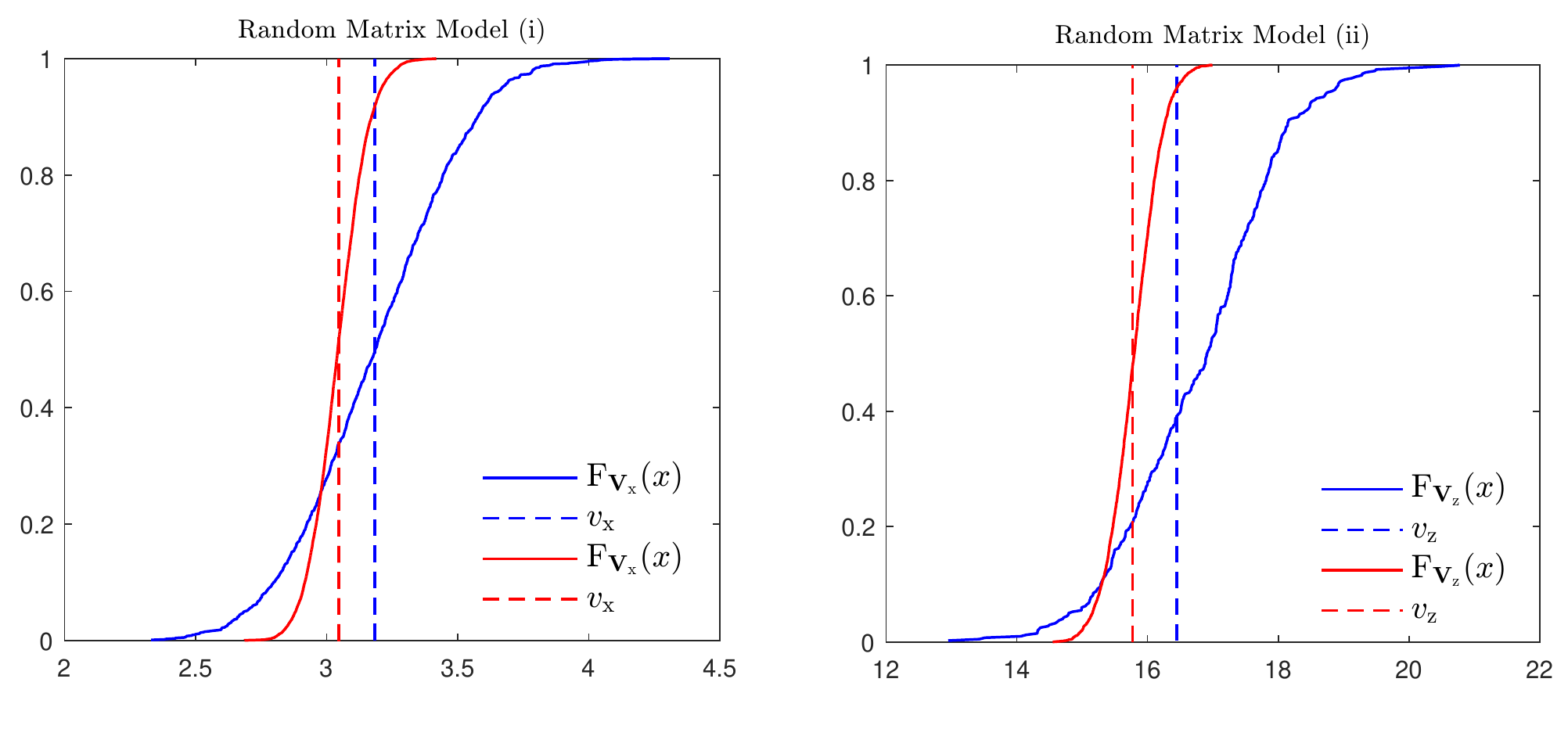,width=2\columnwidth}
	\caption{Empirical cumulative distribution function of the cavity variances. The dimensions of $\matr H$ are ${K/3\times K}$, $\rho=0.1$ and $\tau=1$. Blue curves are for $K=1200$ and red curves are for $K=9600$. The quantities $v_{\rm x}$ and $v_{\rm z}$ are obtained from the stable solutions of self-averaging EP.}
\end{figure*}
\begin{figure*}
	\centering
	\epsfig{file=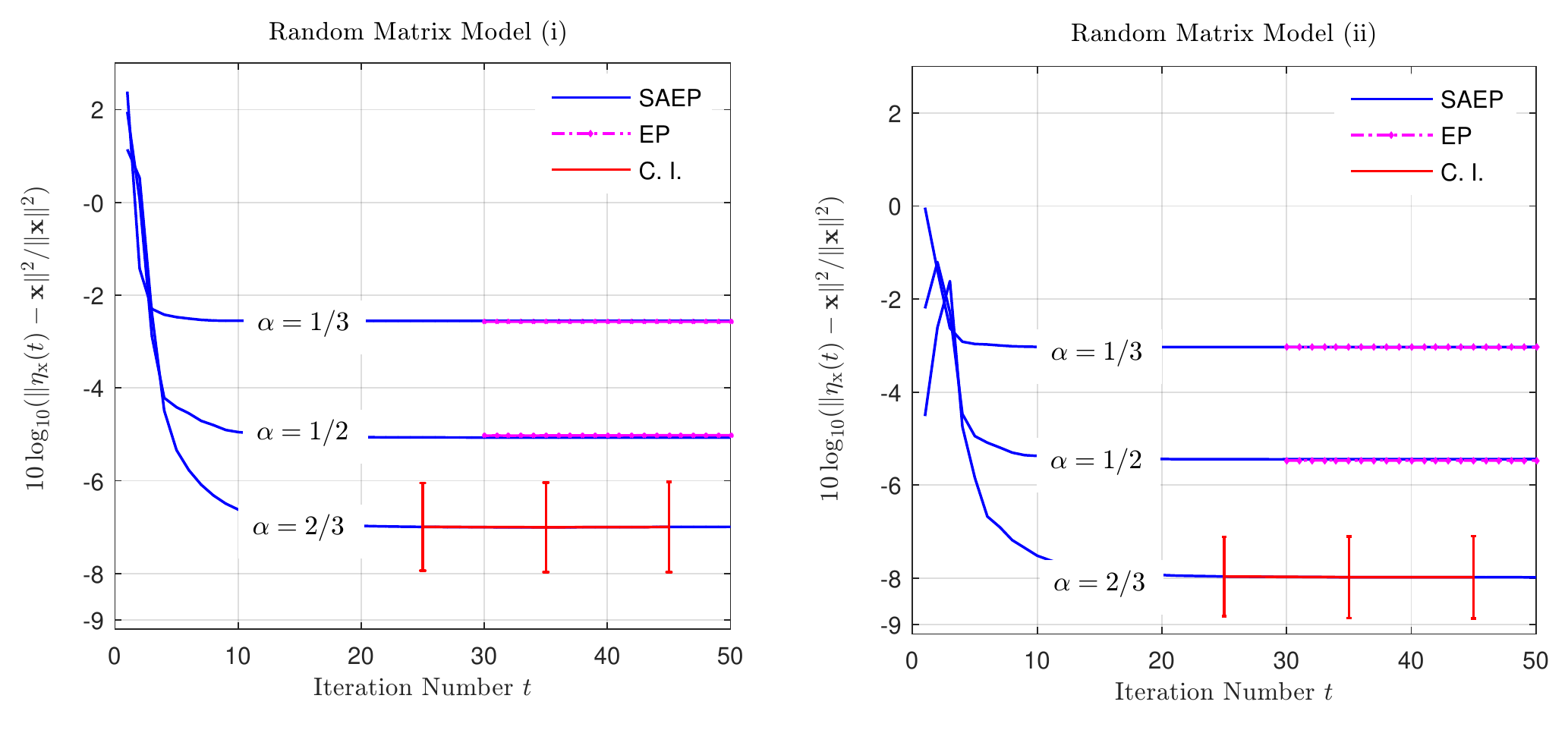,width=2\columnwidth}
	\caption{Mean square error of EP and self-averaging EP (SAEP) versus number of iterations:
		$\matr H $ has dimensions ${\alpha 1200\times 1200}$, $\rho=0.1$ and $\tau=1$. The reported figures are empirical averages over $100$ and $1000$ trials for $\alpha\in\{1/3,1/2\}$ and $\alpha=2/3$ respectively. C.I. denotes the confidence interval in dB.}
\end{figure*}

To solve the EP and self-averaging EP fixed-point equations we consider AMP-like iterative equations. Specifically, the iterative equations \eqref{AMP1}--\eqref{AMPS} are common to EP and self-averaging EP. The iterative equations for the cavity variances $\matr\LLs_{\rm x}(t)$ and $\matr \LLs_{\rm z}(t)$ in EP are readily obtained from \eqref{fix2}, see Appendix~E. The cavity variances in self-averaging EP, i.e. $\matr \LLs_{\rm x}(t)=v_{\rm x}(x){\bf I}$ and $\matr\LLs_{\rm z}(t)=v_{\rm z}(t){\bf I}$, are updated according to 
\begin{align}
v_{\rm z}(t)&= \left(\frac{1}{\langle\matr \chi_{\rm x}(t)\rangle}-v_{\rm x}(t-1)  \right) {\rm S}_{\matr H \matr H^\dagger}\left(-\frac{v_{\rm x}(t-1)\langle\matr \chi_{\rm x}(t)\rangle}{\alpha}\right) \\
v_{\rm x}(t)&=\frac{\alpha (1-v_{\rm z}(t)\langle\matr \chi_{\rm z}(t)\rangle)}{\langle\matr \chi_{\rm x}(t)\rangle}.
\end{align}
For random matrix model (i) we have 
\begin{equation}
{\rm S}_{\matr H\matr H^\dagger}(z)=\frac{1}{1+\alpha z}
\end{equation}
and thereby $v_{\rm z}(t)=1/{\langle\matr \chi_{\rm x}(t)\rangle}$. Hence, we have exactly the AMP algorithm, see \eqref{ampcav}. For random matrix model (ii) we basically have ${\rm S}_{\matr H\matr H^\dagger}(z)=1$ because $\matr H^\dagger \matr H={\bf I}$.

Figure~1 illustrates the convergence of the empirical distributions of the cavity variances i.e. the entries of $\matr \LLs_{\rm x}$ and $\matr \LLs_{\rm z}$, for example
\begin{equation}
{\rm F}_{\matr{\tiny\LL}_{\small\rm x}}(x)=\frac{1}{K}\left\vert\left\{\small\LL \in \{[\small \matr\LL_{\rm x}]_{ii}:\forall i \}: \small\LL\leq x\right\}\right\vert,
\end{equation}
as the dimensions of the system increase. The numerical results show that the cavity variances converge towards their self averaging EP values as the system dimensions increase.

In Figure~2 we compare the performance of EP and self-averaging EP by means of their mean square error in predicting 
the signals. The results show that they both provide the same performance but the latter is outstandingly less complex than the former. 

The provided algorithm for EP is naive and it has a poor convergence because it requires proper iterative scheme for solving a \emph{large number of variables}, e.g. the diagonal entires of $\matr \Lambda_{\rm x}$ and $\matr \Lambda_{\rm z}$, instead of two scalars as in self-averaging EP, e.g. $\lambda_{\rm x}$ and $\lambda_{\rm z}$. One may need to adaptively select the initialization of the EP algorithm to the given parameter values of the system model. On the other hand, self-averaging EP shows excellent convergence for both random matrix models. This can be explained as follows. Recall first that self-averaging EP for random matrix model (i) coincides with AMP, which is known to have excellent convergence. Moreover, we can consider that random matrix model (ii) \emph{asymptotically} behave as an $N\times K$ corner of a (real) Haar matrix with $N,K\to \infty$. It is known that the entries of a Haar matrix have zero mean with variance $1/K$, but they are (weakly) correlated. In other words, random matrix model (ii) has \emph{approximately} similar statistical characteristics as model (i). Figure~2 also confirms this similarity: both algorithms show similar performance for both models. Thereby, it is reasonable to expect that self-averaging EP for matrix model (ii) with iterative equations similar to AMP exhibits a good convergence behavior as well. A similar trick is used in a linear compressed sensing problem \cite{Samp}.

\section{Summary and Outlook} 
For large systems, updating variance parameters of Gaussian EP can require tremendous computational complexity. We have introduced a theoretical framework -- called self-averaging EP -- that transforms this challenge into an opportunity provided the underlying transformation matrix  fulfills certain asymptotic freeness conditions. Self-averaging EP extends previous AMP algorithm -- whose optimality is valid only for the classical iid random matrix ensemble -- to a general class of random matrix ensembles. We have restricted ourselves to cases where the random matrix ensemble is known explicitly. This is typically the case for applications in compressed sensing. But we expect that self-averaging EP can be applied to any model in which latent variables contribute identically (in a statistical sense) to the data such that the random matrix assumption is reasonable. It would then be important to have an estimator of the R-transform (or the S-transform) that is computationally more efficient than the simple one in (\ref{finite1}). This estimator could e.g.\ be based on spectral moments ${\rm Tr}((\matr H^\dagger \matr H)^k)$ up to some order, see \cite{Jolanta}. It would also be interesting to apply methods of random matrix theory to derive convergent algorithms for solving the self-averaging EP fixed point equations, see \cite{Opper16}. 
\section*{Acknowledgment}
The authors would like to thank Yoshiyuki Kabashima for inspiring us to do this work and Mihai Alin Badiu for discussions on the exposition of the paper. This work was partially supported by the research project VIRTUOSO (funded by Intel Mobile Communications, Anite, Telenor, Aalborg University, and the Danish National Advanced Technology Foundation). 
\appendix
\section{Proof of Theorem~1}
\subsection{Preliminaries}
In this section we introduce some preliminary results on the R-transform and the S-transform.
\begin{lemma}\label{Rcar}
	Let ${\rm F}$ be a probability distribution with support in $[0,\infty)$ and its R-transform $\rm  R$. Moreover, let $\chi\triangleq\int x^{-1}{\rm dF}(x)$ where by convention $\chi=\infty$ unless ${\rm F}(0)=0$.  Then, unless ${\rm F}$ is a Dirac distribution, $\rm R$ is strictly increasing on $(-\chi,0)$. In particular, we have $\lim_{\omega\to -\chi^+}{\rm R}(\omega)= \chi^{-1}$ and we have $\chi{\rm R}(-\chi)=1$ unless $\chi=\infty$. 
\end{lemma} 
Lemma~\ref{Rcar} can be easily proved by following the arguments of \cite[pp.446]{Guionnet}.
\begin{lemma}\label{Rlogdet}
	Let ${\rm F}$, $\chi$, $\rm R$ be given as in Lemma~\ref{Rcar}. Furthermore, let $a\in (0,\chi)$ and define the positive quantity
	\begin{equation}
	\epsilon\triangleq \frac{1}{a}-{\rm R}(-a).
	\end{equation}
	Then, we have
	\begin{equation}
	\int_{0}^{a}{\rm R}(-\omega) \;{\rm d}\omega= 1+\ln a-\epsilon a+\int \ln (\epsilon+x)\;{\rm dF}(x).\label{R1}
	\end{equation} 
\end{lemma}
This identity follows by a proper reformulation of \cite[Eq.~(B16)]{Opper16}. Here, the variable $\epsilon$ can be uniquely defined through the implicit equation $a(\epsilon)=\int(\epsilon+x)^{-1}{\rm dF}(x)$. Moreover, $a(\epsilon)\to \chi$ as $\epsilon \to 0$. Thereby, for  $\chi<\infty$ we have
\begin{equation}
\int \ln(x) \;{\rm dF}(x)=-(1+\ln \chi)+\int_{0}^{\chi}{\rm R}(-\omega)\;{\rm d}\omega. \label{R2}
\end{equation}
\begin{lemma}\label{Slogdet}
	Let ${\rm F}$ be a probability distribution with support in $[0,\infty)$ and $\rm S$ its S-transform. Furthermore, let $\alpha\triangleq 1-{\rm F}(0)\neq 0$. Moreover, let $0<  b <\alpha$ and introduce the positive quantity
	\begin{equation}
	\epsilon \triangleq \frac{1-b}{b{\rm S}(-b)}. \label{St}
	\end{equation}
	Then, we have
	\begin{equation}
	\int_{0}^{b}\ln{\rm S}(-z)\;{\rm d}z= H(b)+(1-b)\ln\epsilon-\int \ln(\epsilon+x)\;{\rm dF}(x)\label{S1}
	\end{equation}
	where $H(x)\triangleq(x-1)\ln(1-x)-x\ln x$, i.e. the binary entropy function with natural logarithm.
\end{lemma}
This result is obtained by a convenient reformulation of \cite[Eq.~(23)]{Burak15}. Here, by the definition of the S-transform, the variable $\epsilon$ can be uniquely defined by the implicit equation
\begin{equation}
b(\epsilon) = 1-\int \frac{{\rm dF}(x)}{1+\frac{x}{\epsilon}}
\end{equation}
Thus, for $0<\epsilon<\infty$, we can recast \eqref{S1} as follows
\begin{equation}
\int \ln(\epsilon+x)\;{\rm dF}(x)=H(b)+(1-b)\ln\epsilon-\int_{0}^{b}\ln{\rm S}(-z)\;{\rm d}z \label{S2}
\end{equation}
where for short we wrote $b=b(\epsilon)$. 

\subsection{Derivation of Theorem~\ref{main}}\label{proofmain} 
Consider an $N\times N$ positive definite matrix $\matr X$ with a LED ${\rm F_{\matr X}}$. Let $\phi(\matr X)$ and $\phi(\matr X^{-1})$ be finite. Then, we have \cite[Appendix~C~Proposition.~1]{Burak15}
\begin{equation}
\lim_{N\to \infty}\frac{1}{N}\ln \vert\matr X \vert=\int \ln(x){\rm dF}_{\matr X}(x).\label{conv}
\end{equation}
For instance let $\matr X=\matr \Lambda_{\rm x}+\matr H\matr \Lambda_{\rm z}\matr H$. Since the provided matrices in Theorem~1 are assumed to have compactly supported LEDs, we have  $\phi(\matr X)<\infty$. Moreover, $\phi(\matr X^{-1})<\infty$ is an assumption of Theorem~1. Thus, we have
\begin{equation}
\lim_{K\to \infty}\frac{1}{K}\ln \vert \matr \Lambda_{\rm x}+\matr H^\dagger \matr \Lambda_{\rm z}\matr H\vert=\int \ln(x){\rm dF}_{\matr \Lambda_{\rm x}+\matr H^\dagger \matr \Lambda_{\rm z}\matr H}(x) 
\end{equation}
with the ratio $\alpha=N/K$ fixed. Then, invoking successively \eqref{R2} and additive free convolution \eqref{adf}  we have 
\begin{align}
&\int \ln(x){\rm dF}_{\matr \Lambda_{\rm x}+\matr H^\dagger \matr \Lambda_{\rm z}\matr H}(x)\nonumber \\
&=\int_{0}^{\chi_{\rm x}}{\rm R}_{\matr \Lambda_{\rm x}+\matr H^\dagger \matr \Lambda_z \matr H}(-\omega)\; {\rm d}\omega-\ln \chi_{\rm x}-1\\
&= \int_{0}^{\chi_{\rm x}}{\rm R}_{\matr \Lambda_{\rm x}}(-\omega)\; {\rm d}\omega+\int_{0}^{\chi_{\rm x}}{\rm R}_{\matr H^\dagger \matr \Lambda_z \matr H}(-\omega)\; {\rm d}\omega-\ln \chi_{\rm x}-1\label{adfs}
\end{align}
where $\chi_{\rm x}=\phi(\matr \Lambda_{\rm x}+\matr H^\dagger \matr \Lambda_{\rm z}\matr H)^{-1}$. Here for convenience we introduce 
\begin{align}
v_{\rm x}&\triangleq{\rm R}_{\matr H^\dagger \matr \Lambda_z \matr H}(-\chi_{\rm x})\\
\lambda_{\rm x}&\triangleq{\rm R}_{\matr \Lambda_{\rm x}}(-\chi_{\rm x}).
\end{align}
Then, by invoking Lemma~\ref{Rcar} we can write
$\chi_{\rm x}=1/(\lambda_{\rm x}+v_{\rm x})$. Moreover, with the  conditions stated in Assumption~1 that $\chi_{\rm x}<{\phi}(\matr A^\dagger \matr \Lambda_z \matr A)^{-1}$ and ${\rm F}_{\matr H^\dagger \matr \Lambda_z \matr H}$ is supported on $[0,\infty)$, it turns out that the both quantities  $\lambda_{\rm x}$ are $v_{\rm x}$ positive. At this stage we also note that $\lambda_{\rm x}$ can be interpreted via the "scalarization``
\begin{equation}
{\phi}(\matr \Lambda_{\rm x}+\matr H^\dagger \matr \Lambda_z \matr H)^{-1}= {\phi}(\lambda_{\rm x}\matr{\bf I}+\matr H^\dagger \matr \Lambda_z \matr H)^{-1}.\label{scalarization} 
\end{equation}
Note that the support of ${\rm F}_{\matr \Lambda_{\rm s}}$ is not necessarily in $[0,\infty)$ but it is assumed to be compact. Hence, we can define an auxiliary positive definite matrix $\matr \Lambda_{\epsilon}\triangleq \matr \Lambda_{\rm x}+\epsilon\bf{I}$, such that there exists a positive $\epsilon<\infty$ where $\phi(\matr \Lambda_{\epsilon}^{-1})<\chi_{\rm x}$. Doing so we can write 
\begin{equation}
\int_{0}^{\chi_{\rm x}}{\rm R}_{\matr \Lambda_{\rm x}}(-\omega)\; {\rm d}\omega=-\chi_{\rm x}\epsilon +\int_{0}^{\chi_{\rm x}}{\rm R}_{\matr \Lambda_{\epsilon}}(-\omega)\; {\rm d}\omega.
\end{equation}
In this way both integrals in \eqref{adfs} are recast in a form suitable for applying Lemma~\ref{Rlogdet} from which we easily obtain that 
\begin{align}
&\int \ln(x)\;{\rm dF}_{\matr \Lambda_{\rm x}+\matr H^\dagger \matr \Lambda_{\rm z}\matr H}(x)\nonumber \\
&=\int \ln(x+v_{\rm x})\;{\rm dF}_{\matr \Lambda_{\rm x}}(x)+ \int \ln(x+\lambda_{\rm x})\;{\rm dF}_{\matr H^\dagger \matr \Lambda_{\rm z}\matr H}(x).\label{ex1}
\end{align}
This result still involves difficult terms via the product $\matr H^\dagger \matr \Lambda_{\rm z}\matr H$. We will resolve these difficulties by means of multiplicative free convolution. We first perform a convenient transformation 
\begin{align}
&\int \ln(x+\lambda_{\rm x})\;{\rm dF}_{\matr H^\dagger \matr \Lambda_{\rm z}\matr H}(x)\nonumber \\
&=(1-\alpha)\ln\lambda_{\rm x}+ \alpha\int \ln(x+\lambda_{\rm x})\;{\rm dF}_{\matr H\matr H^\dagger \matr \Lambda_{\rm z}}(x) \label{trs}.
\end{align} 
We simplify the integral in the right hand side of \eqref{trs} by invoking successively \eqref{S2}, multiplicative free convolution \eqref{mdf} and \eqref{S1} as
\begin{align}
&\int \ln(x+\lambda_{\rm x})\;{\rm dF}_{\matr H\matr H^\dagger \matr \Lambda_{\rm z}}(x)\nonumber \\
&= \int_{0}^{\chi}\ln {\rm S}_{\matr H\matr H^\dagger \matr \Lambda_{\rm z}}(-z) \;{\rm d}z+H(\chi)+(1-\chi)\ln \lambda_{\rm x} \\
&=-\int_{0}^{\chi}\ln {\rm S}_{\matr H\matr H^\dagger}(-z)\; {\rm d}z-\int_{0}^{\chi}\ln {\rm S}_{\matr \Lambda_{\rm z}}(-z)\; {\rm d}z \nonumber \\
&\quad +H(\chi)+(1-\chi)\ln \lambda_{\rm x}\\
&= \int \ln(x+v_{\rm a})\;{\rm dF}_{\matr H\matr H^\dagger}(x)+ \int \ln(x+v_{\rm z})\;{\rm dF}_{\matr \Lambda_{\rm z}}(x)\nonumber \\
&\quad+(1-\chi)\ln \frac{\lambda_{\rm x}}{{v}_{\rm a} {v}_{\rm z}}- H({\chi}).\label{sS}
\end{align}
Here, we define
\begin{align}
\chi &\triangleq{\phi}\left\{{\bf I}-\left(\matr{\bf I}+ \frac{1}{\lambda_{\rm x}}\matr H\matr H^\dagger \matr \Lambda_{\rm z}\right)^{-1} \right\}.\label{tildec}
\end{align}
Moreover, from \eqref{St} we write 
\begin{align}
\lambda_{\rm x}&=\frac{1-\chi}{\chi{\rm S}_{\matr H \matr H^\dagger} (-\chi){\rm S}_{ \matr \Lambda_{\rm z}} (-\chi)}\\
{v}_{\rm a}&= \frac{1-\chi}{\chi{\rm S}_{\matr H \matr H^\dagger}(-\chi)} \label{Va}\\
{v}_{\rm z}&= \frac{1-\chi}{\chi{\rm S}_{\matr \Lambda_{\rm z}}(-\chi)}\label{vzz}.
\end{align}
Thus, we get the identity
\begin{equation}
\lambda_{\rm x}= \frac{\chi }{1-\chi}{v}_{\rm a}{v}_{\rm z} \label{keyy}
\end{equation}
and thereby $(1-\chi)\ln \frac{\lambda_{\rm x}}{{v}_{\rm a} {v}_{\rm z}}- H\left(\chi\right)= \ln \chi$. Combining this with \eqref{sS} and \eqref{trs} yields
\begin{align}
&\int \ln(x+\lambda_{\rm x})\; {\rm dF}_{\matr H^\dagger \matr \Lambda_{\rm z}\matr H}(x)=\int \ln(x+v_{\rm a})\;{\rm dF}_{\matr H^\dagger\matr H}(x)\nonumber \\ &+ \int \ln(x+v_{\rm z})\;{\rm dF}_{\matr \Lambda_{\rm z}}(x)+\ln \chi +(\alpha^{-1}-1) \ln \frac{\lambda_{\rm x}}{v_{\rm s}}\label{ex2}
\end{align}
where we use the transformation 
\begin{align}
\int \ln(x+v_{\rm a})\;{\rm dF}_{\matr H\matr H^\dagger}(x)&=\int \ln(x+v_{\rm a})\;{\rm dF}_{\matr H^\dagger\matr H}(x)\nonumber \\
&+(1-\alpha^{-1})\ln v_{\rm a}.
\end{align}
Now we highlight the following facts:
\begin{itemize}
	\item From \eqref{Va} and \eqref{keyy} we have 
	\begin{align}
	{v}_{\rm z}&= \lambda_{\rm x}{\rm S}_{\matr H \matr H^\dagger}(-\chi). 
	\end{align}
	\item From \eqref{scalarization} and \eqref{tildec} we have 
	\begin{align}
	\chi&=\alpha^{-1}{\phi}\left\{{\bf I}-\left(\matr{\bf I}+\frac{1}{\lambda_{\rm x}}\matr H^\dagger \matr \Lambda_{\rm z}\matr H\right)^{-1} \right\}\\
	&=\alpha^{-1}{\phi}\left\{{\bf I}-\lambda_{\rm x}\left(\lambda_{\rm x}\matr{\bf I}+\matr H^\dagger \matr \Lambda_{\rm z}\matr H\right)^{-1} \right\}\\
	&=\alpha^{-1} {v}_{\rm x}\chi_{\rm x}. 
	\end{align}
	\item By the scaling property of the R-transform (see \cite{ralfc}) we have
	\begin{equation}
	{v}_{\rm x}={\rm R}_{\matr H^\dagger \matr \Lambda_{\rm z}\matr H}(-\chi_{\rm x}) \iff 1={\rm R}_{\frac{1}{{v}_{\rm x}}\matr H^\dagger \matr \Lambda_{\rm z}\matr H}(-{v}_{\rm x} \chi_{\rm x}).
	\end{equation}
	We express this identity in terms of the S-transform via \eqref{trans}. Doing so yields
	\begin{align}
	1&= {\rm S}_{\frac{1}{{v}_{\rm x}}\matr H^\dagger \matr \Lambda_{\rm z}\matr H}(-{ v}_{\rm x}\chi_{\rm x})\\
	&={\rm S}_{\frac{1}{{v}_{\rm x}}\matr H^\dagger \matr H}(-{v}_{\rm x} \chi_{\rm x}){\rm S}_{\matr \Lambda_{\rm z}}(-{v}_{\rm x}\chi_{\rm x}/\alpha)\\
	&=\frac{1}{\lambda_{\rm z}} {\rm S}_{\frac{1}{{v}_{\rm x}}\matr H^\dagger \matr H}(-{ v}_{\rm x}\chi_{\rm x})\\
	&={\rm S}_{\frac{\lambda_{\rm z}}{{v}_{\rm x}}\matr H^\dagger \matr H}(-{v}_{\rm x}\chi_{\rm x})\label{Sscale}
	\end{align}
	where we define $\lambda_{\rm z}\triangleq1/{\rm S}_{\matr \Lambda_{\rm z}}(-{v}_{\rm  x} \chi_{\rm x}/\alpha)$ and \eqref{Sscale} follows from the scaling property of the S-transform, see \cite{ralfc}. We now obtain the scaling factor $\lambda_{\rm z}$ and express the result in terms of the R-transform. Solving for ${v}_{\rm x}$ yields
	\begin{equation}
	{v}_{\rm x}={\rm R}_{\lambda_{\rm z}\matr H^\dagger\matr H}(-\chi_{\rm x}).   
	\end{equation}
	Moreover, note that $\chi={v}_{\rm x}\chi_{\rm x}/\alpha$. Thus, from \eqref{vzz} we have
	\begin{align}
	\lambda_{\rm z}&=1/{\rm S}_{\matr \Lambda_{\rm z}}(-\chi)= \frac{{v}_{\rm z} \chi}{1-\chi}. \label{lambdaz} 
	\end{align}
	\item From \eqref{keyy} we have ${v}_{a}=\lambda_{\rm x}/\lambda_{\rm z}$.
	\item We introduce the auxiliary variable $\chi_{\rm z}=\phi(\matr \Lambda_{\rm z}+ v_{\rm z}{\bf I})$. Then, it follows by definition of the S-transform and from \eqref{vzz} that 
	$\chi=\lambda_{\rm z}\chi_{\rm z}$ with noting the identity \eqref{lambdaz}.
\end{itemize}
By invoking these facts and \eqref{ex2}, we finally obtain that
\begin{align}
&\lim_{K\to \infty}\frac{1}{K}\ln \vert \matr \Lambda_{\rm x}+\matr H^\dagger \matr \Lambda_{\rm z}\matr H\vert=\int \ln(x+v_{\rm x}){\rm dF}_{\matr \Lambda_{\rm x}}(x)\nonumber \\
&+\alpha \int \ln(x+v_{\rm z}){\rm dF}_{\matr \Lambda_{\rm z}}(x)+\nonumber \\
&+\int\ln (\lambda_{\rm x}+\lambda_{\rm z}x){\rm dF}_{\matr H^\dagger \matr H}(x)+\ln\chi_{\rm x}+\alpha\ln\chi_{\rm z} \label{son}
\end{align}
where $\chi_{\rm z}= {\phi}(\matr \Lambda_{\rm z}+ {v}_{\rm z}\matr{\bf I})^{-1}$, $\lambda_{\rm z}=\chi_{\rm z}^{-1}-{v}_{\rm z}$, ${v}_{\rm x} =\lambda_{\rm z}{\rm R}_{\matr H^\dagger\matr H}(-\lambda_{\rm z}\chi_{\rm x})$ and ${v}_{\rm z} =\lambda_{\rm x}{\rm S}_{\matr H\matr H^\dagger}(-\lambda_{\rm z}\chi_{\rm z})$. Here, we note that the R- and S-transforms are continuous functions on their definition domains. Since, $\matr \Lambda_{\rm x}$ and $\matr \Lambda_z$ have a compactly supported LEDs, $\phi(\matr \Lambda_{\rm x}+ {v}_{\rm x}\matr{\bf I})$ and $ {\phi}(\matr \Lambda_{\rm z}+ { v}_{\rm z}\matr{\bf I})$ are both finite. Moreover, $\chi_{\rm x}$ and $\chi_{\rm z}$ are finite as well. Here, we also note that the R- and S-transforms are continuous functions on their definition domains. Moreover because we assume that $\matr H\matr H^\dagger$ has a compactly supported LED with the maximum eigenvalue of $\matr H^\dagger\matr H$ converges to a finite limit in the large system limit, we have ${\rm R}^K_{\matr H^\dagger\matr H}(\omega)\to  {\rm R}_{\matr H^\dagger\matr H}(\omega)$ as $K\to \infty$, see \cite[Lemma~3.3.4]{Hiai}. This property turns out be sufficient for representing a quantity in terms of a finite size representation of the S-transform. In summary, from Proposition at \cite[Appendix~C, Proposition.~1]{Burak15} for sufficiently large $N,K$ we can consider a finite size representation of \eqref{son} via the convenient three $\log\det$ terms in a way that each converges to one of the integrals in \eqref{son}. In this way we obtain Theorem~\ref{main}.

\subsection{The AT Line of Stabilities}\label{SecAT}
The Almedia Thouless (AT) line of stability \cite{AT} is a fundamental concept in spin glass theory. It determines a region where TAP approach or the so-called replica ansatz can provide valid (physical) results \cite{Tanaka2}. For example, it is known that below the AT line of stability, the Replica ansatz predicts the entropy as a negative quantity for the standard Ising model \cite{SK}, \cite{Mezard}. In this section we extend the concept of AT line of stability to our TAP approach. Basically, we will introduce the AT line of stability conditions that dictate that the two matrices $(\matr \Lambda_{\rm x}+\matr H^\dagger \matr \Lambda_z \matr H)$ and $\matr \Lambda_{\rm z}$ are positive definite. 

In order to derive the AT line of stabilities we follow the conventional approach \cite{Bray} \cite{Hertz} which is investigating the divergence of the so-called total ``susceptibility" 
\begin{align}
\chi_{{\rm x}}^{(2)}&\triangleq \lim_{K\to \infty}\frac{1}{K}\sum_{n,k} (\matr \Lambda_{\rm x}+\matr H^\dagger \matr \Lambda_z \matr H)^{-1}_{n,k}\\
&=\lim_{K\to \infty}\frac{1}{K}\sum_{k}(\matr \Lambda_{\rm x}+\matr H^\dagger \matr \Lambda_z \matr H)^{-2}_{k}.
\end{align}
Under Assumption~\ref{as1} we show in Appendix~\ref{ATproof} that
\begin{equation}
\chi_{{\rm x}}^{(2)}=\frac{\alpha_{\rm x}}{1-\alpha_{\rm x}{\rm R}'_{\lambda_{\rm z}\matr H^\dagger\matr H}(-\chi_{\rm x})}. \label{Dev1}
\end{equation}
Here we have the usual self-averaging EP quantities in the asymptotic form, i.e. $\chi_{\rm a}=\phi \left(\matr \Lambda_{\rm a}+{v}_{\rm a}{\bf I}\right)^{-1}$, $\lambda_{\rm a}=1/\chi_{\rm a}-v_{\rm a}$ for ${\rm a}\in \{{\rm x}, {\rm z} \}$ where ${v}_{\rm a}$ is given of the form \eqref{cavn1} and \eqref{cavn2} with the replacement $\langle \matr \chi_{\rm a} \rangle \to  \chi_{\rm a}$. Moreover, $\alpha_{\rm x}=\phi \left(\matr \Lambda_{\rm a}+{v}_{\rm a}{\bf I}\right)^{-2}$. Here we point out that from the analysis of self-averaging EP in Section~\ref{SEP} we can consider the approximation $\alpha_{\rm x}\simeq \matr \chi_{\rm x}^\dagger \matr \chi_{\rm x}/K$ where $\matr \chi_{\rm x}$ is the variance of the pdf $\tilde q_{\rm x}$ in \eqref{qx}. The total susceptibility $\chi_{{\rm x}}^{(2)}$ diverges at
\begin{equation}
1-\alpha_{\rm x}{\rm R}'_{\lambda_{\rm z}\matr H^\dagger\matr H}(-\chi_{\rm x})= 0.  \label{AT1}
\end{equation}
This is the AT line of stability which dictates that the matrix $(\matr \Lambda_{\rm x}+\matr H^\dagger \matr \Lambda_z \matr H)$ is positive definite. It extends the previous AT line of stability results, e.g. \cite{Adatap} to the general model \eqref{joint}.

The positive definiteness of the matrix $(\matr \Lambda_{\rm x}+\matr H^\dagger \matr \Lambda_z \matr H)$ results in our derivation that the matrix $(\matr \Lambda_z^{-1} +\frac{1}{\lambda_{\rm x}}\matr H\matr H^\dagger)\matr \Lambda_{\rm z}$ is positive definite too (asymptotically). Therefore, $\matr \Lambda_{\rm z}$ is positive definite if $(\matr \Lambda_z^{-1} +\frac{1}{\lambda_{\rm x}}\matr H\matr H^\dagger)$ is positive definite. Hence, we now introduce the total ``susceptibility" as
\begin{equation}
\chi_{{\rm z}}^{(2)}\triangleq \lim_{N\to \infty}\frac{1}{N}\sum_{n,k}(\matr \Lambda_z^{-1} +\frac{1}{\lambda_{\rm x}}\matr H\matr H^\dagger)^{-1}_{n,k}.
\end{equation}
To calculate $\chi_{{\rm z}}^{(2)}$ in addition to Assumption~\ref{as1} we assume that $\matr \Lambda_{\rm z}^{-1}$ and $\matr H \matr H^\dagger$ are asymptotically free. This is indeed a mild assumption due to \eqref{mdf}. Following the derivation of \eqref{Dev1} in Appendix~\ref{ATproof} one can show that 
\begin{equation}
\chi_{{\rm z}}^{(2)}=\frac{\alpha_{\rm m}}{1-\alpha_{\rm m}{\rm R}'_{\frac{1}{\lambda_{\rm x}}\matr H\matr H^\dagger}(-\chi_{\rm m})}. 
\end{equation}
Here we define $\chi_{\rm m}\triangleq v_{\rm z}(1-v_{\rm z}\chi_{\rm z})$ and $\alpha_{\rm m}\triangleq \phi \left(\matr \Lambda_{\rm z}^{-1}+{v}_{\rm z}^{-1}{\bf I}\right)^{-2}$. Moreover, from the analysis of self-averaging EP in Section~\ref{SEP} we can consider the approximation 
\begin{equation}
\alpha_{\rm m}\simeq \frac{1}{N}\sum_{i}(v_{\rm z}(1-v_{\rm z}[\matr \chi_{\rm z}]_i))^2
\end{equation}
where $\matr \chi_{\rm z}$ is the variance of  the pdf $\tilde q_{\rm z}$ in \eqref{qz}. The total susceptibility $\chi_{{\rm z}}^{(2)}$ diverges at
\begin{equation}
1-\alpha_{\rm m}{\rm R}'_{\frac{1}{\lambda_{\rm x}}\matr H\matr H^\dagger}(-\chi_{\rm m})=0. \label{Dev2}
\end{equation}
We thereby complete the AT line of analysis for the positive definiteness of $(\matr \Lambda_{\rm x}+\matr H^\dagger \matr \Lambda_z \matr H)$ which is \eqref{Dev1}, and the positive definiteness of $\matr \Lambda_{\rm z}$ which is \eqref{Dev1} and \eqref{Dev2}. 

\subsection{Derivation of Equation~\eqref{Dev1}}\label{ATproof}
In order to calculate the susceptibility $\chi_{{\rm x}}^{(2)}$ we introduce 
\begin{equation*}
\chi_{\rm x}(\omega)\triangleq \phi (\matr \Lambda_{\rm x}+\matr H^\dagger \matr \Lambda_z \matr H-\omega {\bf I})^{-1}, \quad \omega\in(-\infty,0).
\end{equation*}
Then, we have
\begin{align}
&\chi_{{\rm x}}^{(2)}= \lim_{\omega\to 0}\frac{\partial}{\partial \omega}\chi_{{\rm x}}(\omega)  \\
&= \lim_{\omega\to 0}\frac{\partial}{\partial \omega} \phi (\matr \Lambda_{\rm x}+ \{{\rm R}_{\matr H^\dagger \matr \Lambda_z \matr H}(-\chi_{\rm x}(\omega))-\omega\} {\bf I})^{-1} \label{adf1}\\
&=\phi \left(\matr \Lambda_{\rm x}+ {\rm R}_{\matr H^\dagger \matr \Lambda_z \matr H}(-\chi_{\rm x}){\bf I}\right)^{-2}\left[1+\chi'_{\rm x}(0){\rm R}'_{\matr H^\dagger \matr \Lambda_z \matr H}(-\chi_{\rm x})\right] \\
&=\phi \left(\matr \Lambda_{\rm x}+ {\rm R}_{\lambda_{\rm z}\matr H^\dagger\matr H}(-\chi_{\rm x}){\bf I}\right)^{-2}\left[1+ \chi'_{\rm x}(0){\rm R}'_{\lambda_{\rm z}\matr H^\dagger\matr H}(-\chi_{\rm x})\right]\label{mdf1}
\end{align}
where for short we wrote $\chi_{\rm x}=\chi_{\rm x}(0)$ and $\chi'_{\rm x}(\omega)=\partial\chi_{\rm x}(\omega)/\partial \omega$. The results \eqref{adf1} and \eqref{mdf1} are the consequences of \eqref{adf} and \eqref{mdf}, respectively. With a convenient reformulation of the right-hand side of \eqref{mdf1} one can conclude \eqref{Dev1}. 
\subsection{Updating EP Cavity Variances} 
The update equations \eqref{AMP1}--\eqref{AMPS} are common to EP and self-averaging EP. The update equations for the cavity variances $\matr\LLs_{\rm x}(t)$ and $\matr \LLs_{\rm z}(t)$ in EP are obtained from \eqref{fix2} as
\begin{align}
[\matr\Lambda_{\rm z}(t)]_{ii}&= 1/[\matr\chi_{\rm z}(t-1)]_{i}- [\matr\LLs_{\rm z}(t-1)]_{ii}\\
\matr\Sigma_{\rm x}(t)&=(\matr \Lambda_{\rm x}(t-1)+\matr H^\dagger \matr \Lambda_{\rm z}(t)\matr H)^{-1}\\
[\matr \LLs_{\rm z}(t)]_{ii}&= 1/[(\matr H \matr \Sigma_{\rm x}(t)\matr H^\dagger)]_{ii}- [\matr \Lambda_{\rm z}(t)]_{ii}\\
[\matr\Lambda_{\rm x}(t)]_{ii}&=  1/[\matr\chi_{\rm x}(t)]_{i}-[\matr\LLs_{\rm x}(t-1)]_{ii}\\
[\matr \LLs_{\rm x}(t)]_{ii}&= 1/[(\matr \Sigma_{\rm x}(t))]_{ii}- [\matr \Lambda_{\rm x}(t)]_{ii}.
\end{align}
The provided EP algorithm is naive and it has a poor convergence, in particular for large $\alpha=N/K$. One may need to adaptively select the initialization of the algorithm to the given parameter values.

\bibliographystyle{IEEEtran}
\bibliography{liter}

\begin{thebibliography}{10}
\providecommand{\url}[1]{#1}
\csname url@samestyle\endcsname
\providecommand{\newblock}{\relax}
\providecommand{\bibinfo}[2]{#2}
\providecommand{\BIBentrySTDinterwordspacing}{\spaceskip=0pt\relax}
\providecommand{\BIBentryALTinterwordstretchfactor}{4}
\providecommand{\BIBentryALTinterwordspacing}{\spaceskip=\fontdimen2\font plus
\BIBentryALTinterwordstretchfactor\fontdimen3\font minus
  \fontdimen4\font\relax}
\providecommand{\BIBforeignlanguage}[2]{{%
\expandafter\ifx\csname l@#1\endcsname\relax
\typeout{** WARNING: IEEEtran.bst: No hyphenation pattern has been}%
\typeout{** loaded for the language `#1'. Using the pattern for}%
\typeout{** the default language instead.}%
\else
\language=\csname l@#1\endcsname
\fi
#2}}
\providecommand{\BIBdecl}{\relax}
\BIBdecl

\bibitem{Minka1}
T.~P. Minka, ``Expectation propagation for approximate bayesian inference,'' in
  \emph{Proceedings of the 17th Conference in Uncertainty in Artificial
  Intelligence}, ser. UAI '01, 2001, pp. 362--369.

\bibitem{OW0}
M.~Opper and O.~Winther, ``Gaussian processes for classification: Mean-field
  algorithms,'' \emph{Neural Computation}, pp. 2655--2684, 2000.

\bibitem{Opper:2013}
M.~Opper, U.~Paquet, and O.~Winther, ``Perturbative corrections for approximate
  inference in gaussian latent variable models,'' \emph{J. Mach. Learn. Res.},
  vol.~14, no.~1, pp. 2857--2898, jan 2013.

\bibitem{Kabashima}
Y.~Kabashima, ``{A CDMA multiuser detection algorithm on the basis of belief
  propagation},'' \emph{Journal of Physics A: Mathematical and General}, vol.
  36.43, p. 11111, October 2003.

\bibitem{Donoha}
D.~L. Donoho, A.~Maleki, and A.~Montanari, ``{Message-passing algorithms for
  compressed sensing},'' \emph{Proceedings of the National Academy of
  Sciences}, vol. 106, pp. 18\,914--18\,919, September 2009.

\bibitem{Rangan}
S.~Rangan, ``Generalized approximate message passing for estimation with random
  linear mixing,'' in \emph{Proc. IEEE International Symposium on Information
  Theory (ISIT)}, Saint-Petersburg, Russia, July 2011.

\bibitem{Adatap}
M.~Opper and O.~Winther, ``{Adaptive and self-averaging
  Thouless-Anderson-Palmer mean field theory for probabilistic modeling},''
  \emph{Physical Review E}, vol.~64, pp. 056\,131--(1--14), October 2001.

\bibitem{Kab08}
Y.~Kabashima, ``Inference from correlated patterns: a unified theory for
  perceptron learning and linear vector channels,'' \emph{Journal of Physics:
  Conference Series}, vol.~95, no.~1, 2008.

\bibitem{OW5}
M.~Opper and O.~Winther, ``Expectation consistent approximate inference,''
  \emph{Journal of Machine Learning Research}, 6 (2005): 2177-2204.

\bibitem{Florent}
F.~Krzakala, M.~M\'{e}zard, F.~Sausset, Y.~Sun, and L.~Zdeborov\'{a},
  ``Probabilistic reconstruction in compressed sensing: algorithms, phase
  diagrams, and threshold achieving matrices,'' \emph{Journal of Statistical
  Mechanics: Theory and Experiment}, vol. 2012, no. P08009, 2012.

\bibitem{Hiai}
F.~Hiai and D.~Petz, \emph{The Semicirle Law, {F}ree {R}andom {V}ariables and
  {E}ntropy}.\hskip 1em plus 0.5em minus 0.4em\relax American Mathematical
  Society, 2006.

\bibitem{Sethna}
J.~Sethna, \emph{Statistical mechanics: entropy, order parameters, and
  complexity}.\hskip 1em plus 0.5em minus 0.4em\relax Oxford University Press,
  2006, vol.~14.

\bibitem{mehta}
M.~L. Mehta, \emph{{Random matrices}}.\hskip 1em plus 0.5em minus 0.4em\relax
  {Elsevier Academic press}, 2004.

\bibitem{Greg}
G.~W. Anderson and B.~Farrell, ``Asymptotically liberating sequences of random
  unitary matrices,'' \emph{Advances in Mathematics}, vol. 255, pp. 381 -- 413,
  2014.

\bibitem{SAMPI}
B.~\c{C}akmak, O.~Winther, and B.~H. Fleury, ``S-amp for non-linear observation
  models,'' in \emph{2015 IEEE International Symposium on Information Theory
  (ISIT)}, June 2015, pp. 2807--2811.

\bibitem{MCB}
K.~B. Petersen and M.~S. Pedersen, \emph{{The Matrix Cookbook}}.\hskip 1em plus
  0.5em minus 0.4em\relax Copenhagen, Denmark: Technical University of Denmark,
  2008.

\bibitem{oxfordSpeicher}
G.~Akemann, J.~Baik, and P.~Di~Francesco, Eds., \emph{{The Oxford Handbook of
  Random Matrix Theory}}.\hskip 1em plus 0.5em minus 0.4em\relax Oxford
  University Press, 2011.

\bibitem{Enzo}
E.~Marinari, G.~Parisi, and F.~Ritort, ``{Replica field theory for
  deterministic models. ii. a non-random spin glass with glassy behaviour},''
  \emph{Journal of Physics A: Mathematical and General}, vol. 27.23, p. 7647,
  October 1994.

\bibitem{Voi92}
D.~V. Voiculescu, K.~J. Dykema, and A.~Nica, \emph{Free random
  variables}.\hskip 1em plus 0.5em minus 0.4em\relax American Mathematical
  Society, 1992, vol.~1.

\bibitem{hager}
U.~Haagerup and S.~M{\"o}ller, ``{The law of large numbers for the free
  multiplicative convolution},'' \emph{Operator Algebra and Dynamics. Springer
  Proceedings in Mathematics \& Statistics}, vol.~58, pp. 157--186, 2013.

\bibitem{Collins-a}
B.~Collins and P.~Sniady, ``{Integration with respect to the Haar measure on
  unitary, orthogonal and symplectic group},'' \emph{Communications in
  Mathematical Physics}, vol. 264.3, pp. 773--795, 2006.

\bibitem{onebit}
\BIBentryALTinterwordspacing
C.~TEMPLATES. (2012) 1bit compressive sensing. [Online]. Available:
  \url{http://dsp.rice.edu/1bitCS/}
\BIBentrySTDinterwordspacing

\bibitem{Candes}
E.~J. Candes and T.~Tao, ``Near-optimal signal recovery from random
  projections: Universal encoding strategies?'' \emph{IEEE Transactions on
  Information Theory}, vol.~52, no.~12, pp. 5406--5425, Dec 2006.

\bibitem{Mikko}
M.~Vehkapera, Y.~Kabashima, and S.~Chatterjee, ``Analysis of regularized ls
  reconstruction and random matrix ensembles in compressed sensing,''
  \emph{IEEE Transactions on Information Theory}, vol.~62, no.~4, pp.
  2100--2124, 2016.

\bibitem{Tulino13}
A.~M. Tulino, G.~Caire, S.~Shamai, and S.~Verd\'{u}, ``Support recovery with
  sparsely sampled free random matrices,'' \emph{IEEE Trans. Information
  Theory}, vol.~59, pp. 4243--4271, 2013.

\bibitem{Samp}
B.~\c{C}akmak, O.~Winther, and B.~H. Fleury, ``{S-AMP: Approximate message
  passing for general matrix ensembles},'' in \emph{Proc. IEEE Information
  Theory Workshop (ITW)}, Hobart, Tasmanina, Australia, November 2014.

\bibitem{Jolanta}
J.~Pielaszkiewicz, D.~von Rosen, and M.~Singull, ``Cumulant-moment relation in
  free probability theory,'' \emph{Acta et Commentationes Universitatis
  Tartuensis de Mathematica}, vol. 18.2, pp. 265--278, 2014.

\bibitem{Opper16}
M.~Opper, B.~{\c{C}}akmak, and O.~Winther, ``A theory of solving tap equations
  for ising models with general invariant random matrices,'' \emph{Journal of
  Physics A: Mathematical and Theoretical}, vol.~49, no.~11, p. 114002, 2016.

\bibitem{Guionnet}
A.~Guionnet and M.~Ma\"{\i}da, ``{A Fourier view on the {R}-transform and
  related asymptotics of spherical integrals},'' \emph{Journal of Functional
  Analysis}, vol. 222, no.~2, pp. 435 -- 490, 2005.

\bibitem{Burak15}
B.~{\c{C}}akmak, R.~R. M{\"{u}}ller, and B.~H. Fleury, ``Capacity scaling in
  mimo systems with general unitarily invariant random matrices,'' \emph{arXiv
  preprint}, vol. abs/1306.2595v3, December 2015.

\bibitem{ralfc}
R.~R. M{\"u}ller, ``Applications of large random matrices in communications
  engineering,'' in \emph{International Conference on Advances in the Internet,
  Processing, Systems, and Interdisciplinary Research (IPSI)}, Sveti Stefan,
  Montegro, October 2003.

\bibitem{AT}
J.~R.~L. De~Almeida and D.~J. Thouless, ``Stability of the
  sherrington-kirkpatrick solution of a spin glass model,'' \emph{Journal of
  Physics A: Mathematical and General}, vol.~11, no.~5, p. 983, 1978.

\bibitem{Tanaka2}
T.~Tanaka, ``A statistical-mechanics approach to large-system analysis of cdma
  multiuser detectors,'' \emph{IEEE Transactions on Information Theory},
  vol.~48, no.~11, pp. 2888--2910, Nov 2002.

\bibitem{SK}
D.~Sherrington and S.~Kirkpatrick, ``{Solvable model of a spin-glass},''
  \emph{Physical Review Letters}, vol. 35.26, p. 1792, 1975.

\bibitem{Mezard}
M.~M\'{e}zard, G.~Parisi, and M.~Virasoro, \emph{Spin Glass Theory and
  Beyond}.\hskip 1em plus 0.5em minus 0.4em\relax World Scientific, 1987,
  vol.~9.

\bibitem{Bray}
A.~J. Bray and M.~A. Moore, ``{Metastable states, internal field distributions
  and magnetic excitations in spin glasses},'' \emph{Journal of Physics C:
  Solid State Physics}, vol.~14, p. 2629, 1981.

\bibitem{Hertz}
K.~H. Fisher and J.~A. Hertz, Eds., \emph{{Spin Glasses}}.\hskip 1em plus 0.5em
  minus 0.4em\relax Cambridge University Press, 1993, vol.~1.

\end{thebibliography}
\end{document}